\documentclass[a4paper,fleqn]{cas-dc}

\usepackage{mathtools}
\usepackage{xspace}
\usepackage{enumitem}
\usepackage{listings}
\usepackage{natbib}
\usepackage{censor}
\usepackage{amsmath,amssymb,amsfonts}
\usepackage{soul}
\usepackage{array}
\usepackage{multirow}
\usepackage{algorithmic}
\usepackage{graphicx}
\usepackage{textcomp}
\usepackage[table]{xcolor}
\usepackage{tikz}
\usetikzlibrary{shapes.geometric, arrows}
\usepackage{todonotes}
\usepackage{hyperref}
\usepackage[T1]{fontenc}
\usepackage{url}
\usepackage[strings]{underscore}
\usepackage{tcolorbox}
\usepackage{flushend}
\newcommand{\HIDE}[1]{}
\newcommand{\NEW}[1]
{{\color{teal}#1}}
\newcommand{\NEWB}[1]
{{\color{teal}#1}}
\renewcommand{\NEW}[1]
{#1}
\renewcommand{\NEWB}[1]
{#1}

\newcommand{\mpt}{Mean Passing Tests\xspace}
\newcommand{\acceptone}{accept@\!1\xspace}
\newcommand{\fp}{False Positive\xspace}
\newcommand{\fn}{False Negative\xspace}
\newcommand{\errorrate}{Error Rate\xspace}
\newcommand{\fpr}{False Positive Rate\xspace}
\newcommand{\fnr}{False Negative Rate\xspace}

\lstdefinestyle{mystyle}{
    commentstyle=\color{cyan},
    keywordstyle=\color{violet}\sf\bf,
    numberstyle=\tiny\color{codegray},
    stringstyle=\color{violet}\sf\bf,
    basicstyle=\scriptsize\sf,
    breakatwhitespace=false,
    breaklines=true,
    captionpos=b,
    keepspaces=true,
    numbers=none,
    numbersep=5pt,
    showspaces=false,
    showstringspaces=false,
    showtabs=false,
    tabsize=1
}

\begin{document}
\let\WriteBookmarks\relax
\def\floatpagepagefraction{1}
\def\textpagefraction{.001}
\shorttitle{Talk is Cheap, Logic is Hard}
\shortauthors{Prasetya et~al.}

\title[mode = title]{Talk is Cheap, Logic is Hard: Benchmarking LLMs on Post-Condition Formalization}
%\tnotemark[1,2]
%\tnotetext[1]{This document is the results of the research project funded by the National Science Foundation.}
%\tnotetext[2]{blablabla}

\author[1]{I.S.W.B. Prasetya}[orcid=0000-0002-3421-4635]
\ead{s.w.b.prasetya@uu.nl}

%\credit{Conceptualization of this study, Methodology, Software}

%\address[1]{, Street 129, 1043 NX Amsterdam, The Netherlands}
\affiliation[1]{organization={Utrecht University}, city={Utrecht}, country={the Netherlands}}

\author[2]{Fitsum Kifetew}[orcid=0000-0003-1860-8666]
\ead{kifetew@fbk.eu}
\author[2]{Davide Prandi}[orcid=0000-0001-9885-6074]
\ead{prandi@fbk.eu}

%\credit{Data curation, Writing - Original draft preparation}

\affiliation[2]{organization={Fondazione Bruno Kessler}, city={Trento},country={Italy}}

%\nonumnote{This note has no numbers. In this work we demonstrate $a_b$ the formation Y\_1 of a new type of polariton on the interface between a cuprous oxide slab and a polystyrene micro-sphere placed on the slab. }

\begin{abstract}
\textbf{Context:} Formal specifications, such as pre- and post-conditions provide a solid basis for performing thorough program verification and validation. However, developers rarely provide such formal specifications, hence if AI could help in constructing them, it would make formal verification possible or at least make automated testing much more effective.
\textbf{Objective:} This paper presents a study on the ability of Large Language Models (LLMs) in generating formal full pre- and post-conditions of a program, given its informal natural language description.
\textbf{Method:} Twenty-four state-of-the-art LLMs were evaluated on a freshly prepared dataset of 40 tasks. The paper investigates specifications of varying difficulty and discusses a set of more refined performance metrics in addition the general accept@1 performance. It also investigates the impact of using automatically generated tests for validation of the solutions proposed by LLMs.
\textbf{Results:} The results of the experiment reveal that, in general LLMs can produce valid pre- and post-conditions based on natural language descriptions of programs. Incorrect solutions from proprietary models are also often near correct. A closer inspection shows that open-source models tend to result in a higher proportion of erroneous results while proprietary models tend to have slightly higher false negative rates. Interestingly, the results also show that augmenting the manually prepared dataset with automatically generated tests leads to the exposure of wrong solutions, which would have otherwise been accepted as correct.
\textbf{Conclusion:} In general, LLMs perform better in formalizing pre-conditions than on post-conditions and proprietary models perform better than open ones. However, none of the LLMs were able to correctly formalize all the tasks in our benchmark. Overall, the effectiveness and reliability of LLMs in formalizing pre- and post-conditions could be greatly improved by using a good test suite that checks the correctness of the LLM generated formalizations.
%
%
% \HIDE{Many Large Language Models (LLMs) have been trained to do programming completion tasks. The skill has been employed in practice to aid programmers in writing programs.
% Recent studies have showed that LLMs are also capable of doing other programming related tasks, such as writing tests, fixing code, and extracting formal specifications.
% This paper investigates the latter skill.
% Developers rarely provide formal specifications, so if an LLM can help in constructing them, this would make formal verification possible, or at least to make automated testing much more effective, beyond simple detection of program crashes.
% %The paper will in particular look into the role of used test suites and the application different prompting approaches, such as chain of thought, towards the quality of the obtained specifications.
% More specifically, the paper presents a study on LLMs' ability to generate machine interpretable pre- and post-conditions of a program, given its informal natural language description.
% Compared to similar studies in the past, this paper covers a broad set of LLMs, including recent open source LLMs, and addresses  the challenge for specifications of different complexity.
% }
\end{abstract}

%
% WP: No graphical-abstract for now
%
%\begin{graphicalabstract}
%\includegraphics{figs/cas-grabs.pdf}
%\end{graphicalabstract}

%
% WP: no highlights too for now
%
%\begin{highlights}
%\item Research highlights item 1
%\item Research highlights item 2
%\item Research highlights item 3
%\end{highlights}


\maketitle

\section{Introduction}\label{sec.intro}

Many Large Language Models (LLMs) are able to generate new content \citep{feuerriegel2024generative} beyond their original training data ({\em generative AI}).
They have amazed many of us in their ability to engage in  conversations and produce coherent answers when asked questions.
Many LLMs have also been trained in doing code completion tasks. The skill has been employed in practice to aid programmers writing code,  e.g. IDEs like PyCharm and Raider have integrated LLM Copilots.
This leads to the hypothesis that perhaps, with the right prompting, LLMs can also be used to aid other programmimg-related tasks.
For example the use of LLMs to write tests~\citep{yang2024evaluation}, inferring test oracles~\citep{mahmud2025augmentest}, and aid debugging \citep{xia2023keep} were recently studied.
Another interesting application of LLMs is to generate the formal specification of a program from its natural language description.

The presence of a formal specification would enable rigorous checking of the correctness of a program, e.g. through symbolic techniques or through extensive automated testing.
Note that without specifications such techniques can only be used to check for general errors (such as crashing) or regression errors.
Unfortunately in practice developers rarely provide formal specifications.  A possible explanation is that developers are not trained in writing them.
Arguably though, the more likely reason is that companies consider the effort to be too much, so  developers are not required or  encouraged to write them. Hence, if LLMs can produce them, or at least aid developers in producing them, it might lower the cost and make the gain attractive.

Let us first argue that a formal specification does not necessarily have to be expressed in (complex) math formulas. It however needs to be {\em interpretable} by a computer, so that it can be used to validate a program.
%Ultimately, this is the feature that is needed most to be able to reliably and feasibly checking a program's correctness with respect to the said specification.
%Without this feature, a manual pen and paper proof would be needed, which is not feasible, and by itself is error prone.
A 'formal specification' can thus be just a Python function, for example, with which we can check if the output of the associated program is correct. From this perspective, constructing specifications in the form of pre- and post-conditions, also known as {\em contract} \citep{meyerDBC92}, can be phrased as a problem of constructing boolean functions $pre$ and $post$
%(a program that returns a value of type $\sf bool$)
given a natural language description. Many LLMs are trained in such a programming task in general, i.e.,  producing a program from a  natural language description. So, in theory producing pre- and post-conditions should not be much different. However, there is a twist. Consider, for example the following description of a program:

%\begin{equation}\label{eq.prgF}
%\begin{array}{c}
%\vspace{2mm}
\begin{quote}
\hspace{-7mm}
\fbox{\parbox{0.9\columnwidth}{\em The program $sortx(s)$ takes a list with no duplicates,
and returns an increasingly sorted copy of $s$.}}
\end{quote}
%\end{array}
%\end{equation}

\noindent
While the text describes the program $sortx$, it does not directly describes boolean functions $pre$ and $post$. So, it is not given that the LLM's programming skill would automatically transfer to extracting and formalizing pre- and post-conditions.

%to extract the pre- and post-conditions an LLM would need some additional training. The good thing is that quite a number of LLMs seem to understand the task, which suggests that they were also trained to do so.

We will distinguish between {\em full} and {\em partial} or incomplete pre-/post-conditions. \NEW{
A {\em full} post-condition is a post-condition that fully captures the correctness of the intended functionality of the program it specifies.
%For example, a full post-condition for the program $sortx$ above could assert, informally, that the returned list should:
%
%\begin{quote}
%\begin{enumerate}
%    \item \label{F.postcond.1} contain as many elements as $s$, and
%    \item contain every element of $s$, and
%    \item it is sorted increasingly\footnote{Given the pre-condition that $s$ contains no duplicate, there is no need to require that the returned list is a permutation of the original $s$.}.
%\end{enumerate}
%\end{quote}
%
Alternatively, one can come up with a post-condition that only {\em partially} captures the intent.
For example, a post-condition that only asserts that the output of $sortx(s)$ should be of the same length as $s$ is partial.
%We call such a post-condition {\em partial} \cite{xxx}.
%More generally, a partial post-condition is a post-condition that is weaker than a full post-condition.
Analogously, a partial pre-condition only partially captures the intended constraint on the program's inputs.
%A pre-condition that is weaker that a full pre-condition is partial.
%
Partial pre- and post-conditions are easier to write, as we do not need to be complete in capturing a program's intent.
Unfortunately, a partial post-condition is also weaker in detecting errors.
For instance the aforementioned example of a partial post-condition for $sortx$ will accept an implementation that sorts the returned list decreasingly (wrong) instead of increasingly.
Analogously, using a partial pre-condition would trigger spurious bug reports during testing or verification. That is, the post-condition reports a violation/error, but this occurs because the program was called with an input that does not satisfy its intended pre-condition. The reported error is therefore not an actual error. }

Full pre- and post-conditions are therefore preferred. On the other hand they are more challenging to formulate, hence if LLMs can help it would considerably reduce the effort required in formalizing contracts.
In this paper we will focus on the generation of {\bf full} pre- and post-conditions.
%\HIDE{
More specifically, we investigate the following research questions:

\begin{itemize}[leftmargin=*]
    \item{\bf RQ1}: {\em how do LLMs perform in generating {\bf full} formal pre- and post-conditions? How good are open source LLMs compared to propriety ones?}

    \item{\bf RQ2}: {\em how does pre-/post-condition difficulty impact LLMs' performance? }

\end{itemize}

\noindent
We present the benchmarking experiment we performed using a newly crafted dataset and involving a variety of state-of-the-art LLMs to answer these questions. \\

    %[we have to come up with the 'types'. I notice that some pre/posts are easily expressible with list comprehension. Some are more convenient to be coded as imperative loops. Programs of type 'parse string of some grammar' are easy to express with context free grammar, but hard to formalize is pred-logic. I notice LLMs have lots of problem with these, probably because it was not trained in the concepts of regular expr and cfgs.]

%    \item \XCOMMENT{dropped}{\bf RQ3}: can we get correct specifications with a suboptimal test suite?Does adding counter-tests help in getting correct specifications?

%   \item \XCOMMENT{dropped}{\bf RQ4}: do prompt reframing and CoT help in getting correct specifications?
%}

\noindent
{\bf Contribution.}
This paper offers a study on the ability of state-of-the-art LLMs to generate formal pre- and post-conditions of a program, given the program's {\em natural language} specification/description.
%By 'formal' we refer to our aforementioned notion of formal.
The study offers the following advancement with respect to previous studies.

\begin{itemize}[leftmargin=*]
  \item Previous studies~\citep{endres2024can,he2024beyond,ma2024specgen} focused on the generation of {\em partial} post-conditions, which are weaker in detecting faults.
  This paper offers insight into LLMs's ability in generating {\em full} pre- and post-conditions.
  %that capture the whole functionality of the target program.

\item \NEW{Aforementioned previous studies focused on the accept@\!1 metric, similar to pass@\!1 metric \citep{chen2021evaluating}. This metric only considers correct solutions. Our paper offers a set of more refined metrics, e.g. indicating if a solution is {\em almost} correct. Arguably, such a solution is still valuable, e.g. developers can manually fix them based on information from failing test cases. This paper furthermore looks into the performance of LLMs across pre-/post-conditions of different difficulties.
Finally, the paper also discusses several selected cases involving language subtleties and the insight learned from them.
}

  \item Our study assesses a comprehensive set of state-of-the-art proprietary and open LLMs, offering a broader insight into their ability in formalizing contracts.

\end{itemize}

\

\noindent
{\bf Paper structure}.
We start by presenting relevant related work in Section~\ref{sec:releated-work}. Section~\ref{sec:prepost} defines pre- and post-conditions, gives  examples of their formalization, and introduces the concept of specification difficulty.
Section~\ref{sec:methodology} describes our benchmarking methodology.
Section~\ref{sec:experiment-design} states the research questions that we investigate and describes the  design of the experiment carried out to address them.
Section~\ref{sec:results} presents the results, and
Section~\ref{sec:discussion} discusses several selected cases.
Treats to validity are discussed in Section~\ref{sec:threats}.
Finally, Section~\ref{sec:concl} concludes the paper and mentions some future work. \\

% \noindent
% {\textbf{Replication package.} A package with all data and scripts is available via Zenodo \url{https://zenodo.org/records/17499683} as supplementary material.
% %It will be made publicly available upon paper acceptance.
% }

\section{Related Work} \label{sec:releated-work}
LLMs' ability to produce specifications, in particular post-conditions, has been studied in a number of recent papers. 
\cite{he2024beyond} 
studied the generation of partial post-conditions of various types (e.g. null-check $retval {\not=} {\sf None}$ or numerical range such as $0 {\leq} retval {<} n$) given a program's natural language description, for Python programs in the EvalPlus dataset \citep{evalplus} (more specifically, its HumanEval subset).
Strong performance was shown, while using only small LLMs (Gemma-7b and Mistral-7B, with about 7 billion parameters, as opposed to GPT 3.5's 175 billion) where 99\% of the generated post-conditions were validated to be sound. But, being partial they are of course incomplete (not able to detect all bugs).
Extraction of full post-conditions was attempted by 
%Endres et al.~
\cite{endres2024can}. Larger LLMs were used (GPT3.5, GPT4, and Starchat). Despite this, for about 39\% of the programs from the HumanEval subset, the LLMs could not construct a correct post-condition. 
Endres et al. prompted the LLMs to produce a post-condition in the form of a {\em single} $\sf assert$, potentially restricting the LLMs.
% In our approach prompt the LLMs differently.

Since LLMs are also trained to comprehend program code, one may contemplate to ask them to extract pre- and post-conditions directly from the code, or fragments of it.
%, or to just give both the documentation and the code to the LLMs. The latter is also studied in \cite{endres2024can}; the results do not show strong evidence that including the code improves the performance.
This was studied by 
%Wen et al.~
\cite{wen2024enchanting} with their AutoSpec tool which uses GPT-3.5 to generate assertions, including post-conditions, for C programs. Symbolic verification is used to remove candidate assertions which are inconsistent with the program.
%Ma et al.~
\cite{ma2024specgen} proposed SpecGen for generating assertions for Java programs. It also uses GPT-3.5 and a symbolic verifier to validate candidates proposed by the LLM. But SpecGen also includes an {\em iterative} part. If a proposed assertion fails, a prompt is generated for the LLM to fix the proposal, along with some feedback distilled from the verification report. %This process is iterated.
SpecGen also has a loop to evolve candidate specifications through direct mutations (without using the LLM). Relatively high success rates were reported (for about 72\% of the programs in their dataset, sound post-conditions were extracted).
The approach may however produce partial post-conditions; the study itself does not include an assessment on the completeness of the obtained post-conditions.
One should also keep in mind that specifications extracted from the code may inherit bugs already in the code.
\NEWB{
\cite{faria2026automatic} investigated Claude 4.5 and GPT5.2 for generating pre-/post-conditions and loop invariants for Dafny~\citep{leino2010dafny} programs. Given a Dafny program, an LLM is asked to add missing specifications, such as pre/post conditions and loop invariants. The Dafny verifier is used for validation.
Their approach adopts an iterative repair prompting where a repair prompt is sent to the LLM if a proposed specification failed the validation. The repair is reported to greatly improve the success rate in obtaining sound specifications. The work presented in this paper assesses LLMs' potential in producing formal specifications from natural language descriptions, differently from \cite{faria2026automatic}  where the LLM works on a Dafny program.
%together with its description.
}

In AutoSpec and SpecGen, GPT-3.5 is asked to produce  specifications in respectively ACSL and JML; these are dedicated specification languages. Many coding LLMs were trained with Python and Java, but it is unknown whether they were also trained with JML or ACSL.  Though much of JML's syntax is similar to Java, it is still a language of its own.
So if GPT-3.5 was not trained with JML, and yet able to produce JML specifications, that would show that LLMs are also capable of cross-lingual generalization. 
\cite{LLMSymboleo} attempted to use GPT-4o to extract Symboleo specifications from natural language descriptions of business contracts. Symboleo is a dedicated specification language. However, its structure and syntax are different from Java or JML.  Unfortunately, in this experiment GPT-4o struggled in producing correct Symboleo fragments.

Further in the past, research in formal specifications generation mainly focused on techniques to infer them from execution traces ({\em dynamic inference}). 
%We can call computer-constructed specifications, including those constructed by LLMs, as {\em synthetic}.
Tools like Daikon \citep{ernst2007daikon} and AutoInfer \citep{wei2011inferring} can derive pre- and post-conditions. 
\HIDE{
'invariants' from such traces. An invariant is a state predicate that always hold at some chosen control flow location in a program. Pre- and post-conditions are instances of invariants. So are loop-invariants, and Daikon can thus also be used to infer them. Daikon can recognize complex invariants (e.g. involving quantifiers), though} 
Only formulas that are instances of pre-defined formula-patterns can be derived. 
Other types of formal specifications can be derived as well, e.g. 
temporal specifications 
%(e.g. asserting that an action $a$ is always followed by $b$) 
\citep{beschastnikh2011leveraging},  
algebraic specifications 
%(e.g. asserting that actions $a$ and $c$ behave commutatively: $ac = ca$) 
\citep{elyasov2013guided}, 
or even a Finite State Machine (FSM) model \citep{mariani2008automated}.
Dynamic inference is  neither sound nor complete, as we can only collect finite number of traces. 
%So, a specification thus obtained may give a false positive (reporting a program as faulty while it is not) as well as a false negative (approving a program as non-faulty, while it is faulty).
Also, if the program has a bug, the bug may manifest in the traces, and thus also in specifications derived.
%This does not mean that these approaches are of no use. 
LLM-based approaches are also not guaranteed to be sound and complete, due to LLMs' probabilistic nature as well as incompleteness and ambiguity in the program description.
The two approaches are complementary, each with strength and weakness. Combining them seems to be an area worth exploring for future work.

\section{Pre- and Post-conditions}
\label{sec:prepost}
A popular way to formally specify the functionality of a program is by using pre- and post-conditions. The theory was proposed by \cite{hoare1969axiomatic}. The idea was later brought into software development by \cite{meyerDBC92}, who then called them {\em program contract}. 
Consider a program $F(x)$ that takes a single parameter $x$ as input and returns a value as output.
$F$ can be specified using a triple $\{ P \} \; F(x)\; \{ Q \}$. The triple is also called Hoare triple \citep{hoare1969axiomatic}. $P$ is called a {\em pre-condition}: a predicate over the $F$'s input, specifying which input values are valid/allowed for invoking $F$. $Q$ is called a {\em post-condition};
a predicate over $F$'s input and output, specifying what $F$ computes, if it is invoked with an input satisfying $P$. For example, the triple below specifies a program that computes the maximum element of a non-empty list of integers.
\begin{equation}\label{spec:findMax}
%\footnotesize
\begin{array}{lc}
P: & \{ \ |a|>0 \ \}  \\ 
   & {\sf findMax}(a{:}list[int])  \\ 
Q: & \{ \ {\sf retval} \in a \wedge 
      (\forall x\in a \bullet {\sf retval} \geq x)\ \} 
\end{array}
\end{equation}
Various specification languages support pre- and post-conditions e.g. OCL, JML, Z, and the contract language of Eiffel.
If present, pre- and post-conditions are very valuable artifacts. 
Having a post-condition allows a program to be symbolically verified. Alternatively, it allows automated testing to be applied, serving as a general test-oracle to check the program's output for {\em any} input instance that an automated testing algorithm produces. 
%The pre-condition is equally important to prevent false positives during verification or testing.

\subsection{Full and partial conditions} \label{sec:complete-partial}

\NEW{The concept of full and partial was mentioned before. A post-condition $Q$ of a program $F(x)$ is full if it fully captures $F$'s intent. To put it differently, $Q$ is full if it accepts any output of a correct implementation of $F$, and rejects any wrong output of any incorrect implementation of $F$. A partial post-condition $Q'$ also accepts all correct outputs of $F$, but it  also accepts some outputs that are actually wrong. So, a partial post-condition is strictly weaker than a full one ($Q' \not\equiv Q \; \wedge \; Q' \Leftarrow Q$). A pre-condition is full if it approves any input that a correct $F$ should be able to accept, and rejects inputs that a correct $F$ are not expected to accept. A partial pre-condition approves all correct inputs, but it also approves some inputs that should actually be rejected.

For example, the pre- and post-conditions in the specification of $\sf findMax$ in (\ref{spec:findMax}) are full. Below is an alternative specification, but its post-condition is partial: %\todo[]{Should we use findMax everywhere? $\rightarrow$ should be replaced now}
\begin{equation}\label{spec:findMax-partial}
%\footnotesize
\{ \ \underbrace{|a|>0}_P \ \} \ \ {\sf findMax}(a{:}list[int]) \ \ 
   \{ \ \underbrace{{\sf retval} \in a}_{Q'}\  \}
\end{equation}
\noindent
The post-condition $Q'$ is simpler than its full counterpart in (\ref{spec:findMax}).
But if an implementation of $\sf findMax$ returns the smallest element, rather than the greatest element, this bug will unfortunately be left undetected by this $Q'$.
%Similarly, a pre-condition $P'$ of $F(x)$ is partial if the implication $P \Rightarrow P'$ is valid, where $P$ is a full pre-condition of $F(x)$. As an example the precondition $true$ is partial for the program $\sf findMax$ above. When used for testing, this precondition would allow an empty $s$ to be used as a test input, while this is actually not allowed by the intended pre-condition. If the program subsequently crashes, this is not a real error. 
That is why using a partial post-condition, or a partial pre-condition, for testing is not preferred. }

\subsection{Interpretable specification}

To be used for verification or testing, a formal specification has to be {\em interpretable} by a computer. We do not necessarily need to use a dedicated specification language such as OCL or Z to achieve interpretability. We can also write pre- and post-conditions as expressions or code in the programming language of the specified program itself. Pre- and post-conditions are predicates, so they can be written as boolean functions. For example $\sf findMax$'s pre- and post-conditions in (\ref{spec:findMax}) can be written as the following functions in Python:
\begin{equation}\label{eq.pre-post-findmax}
\small
\begin{array}{l}
{\color{violet}\bf def}\;{\sf preFindMax}(a) : {\color{cyan}\bf return}\ {\sf len}(a) > 0 \\
{\color{violet}\bf def}\; {\sf postFindMax}(retval,a)  :  \\
   \hspace{5.5mm} 
   {\color{cyan}\bf return}\  retval\;{\sf in}\; a
   \ {\sf and}\ {\sf all}(retval \geq x \;{\bf for} \;x\; {\bf in}\; a) 
\end{array}
\end{equation}
The paramater $retval$ in the post-condition represents the return value (output) of $\sf findMax$.

For verification or testing we can formulate the original Hoare triple as a function, e.g. as shown below, which we then use as the verification or testing entry point.
\begin{equation}
\small
\begin{array}{l}
{\color{violet}\bf def} \; {\sf verifyFindMax}(a): \\
\hspace{3mm} {\bf if}\; {\sf prefindMax}(a): \\
\hspace{6mm} retval = {\sf findMax}(a) \\
\hspace{6mm}
      {\color{cyan}\bf assert}\; {\sf postfindMax}(retval,a)
\end{array}    
\end{equation}

\subsection{Specification difficulty} \label{sec.spec.diff}

The pre-condition of $\sf findMax$ is just a simple formula $|a|{>}0$. Such a formula should be easy for a programmer to figure out.
However, 
formulas that involve quantifiers, such as the post-condition of $\sf findMax$, are  more challenging to write for humans \citep{prasetya2020can,Bravo23}. Formulas involving nested quantifiers are even more challenging \citep{Bravo23}. We therefore would want to know whether LLMs can help us in formalizing the more difficult specifications.
We define the following classes of \NEW{semantical} difficulty for post-conditions, ordered increasingly in their difficulty. Analogously, we have the same classification for pre-conditions. 

\begin{itemize}[leftmargin=*]
    \item Category $\sf S$: the post-condition is a simple expression. It does not contain any quantified formula. Example: $x{>}0 \wedge len(a){>}0$.
    
    \item Category $\sf Q$: the post-condition has one or more quantified formulas, but they are {\em not nested}. Also, no explicit loop nor recursion is used.
    A list or set comprehension is treated as a quantified expression. Example: 
    $(\forall x: x\in a: x{>}0)$. In Python this is expressible with a list or set comprehension:
    ${\sf all}(x{>}0\; {\bf for}\;x\;{\bf in}\; a)$.
        
    \item Category $\sf QQ$: the post-condition has at least one {\em nested} quantified formula. No explicit loop nor recursion is used.
    Example: $(\forall x: x\in a: (\forall y: y\in b: x{>}y))$. In Python this is expressible with a nested list or set comprehension:
    ${\sf all}(x{>}y\; {\bf for}\;x\;{\bf in}\; a\; {\bf for}\;y\;{\bf in}\; b)$. 
    
    \item Category $\sf NQ$: the post condition uses uses one or more explicit loops or recursions. A loop is defined as explicit if it is a $\sf while$-loop or is a $\sf for$-loop but is not a comprehension. An example of this category is given below.

\end{itemize}

The categories above are disjoint and complete. That is, a given formula always belongs to one of the categories, and only to one of them.

\HIDE{
To later be able to express a notion of {\em average} difficulty, we define the {\em difficulty level} ($di\!f$): $di\!f(\sf S) = 1$, $di\!f(\sf Q) = 2$, $di\!f(\sf QQ) = 4$, and $di\!f(\sf NQ) = 8$. So, each class is considered to be twice more difficult than the class below it.
}

Not all post-conditions are expressible with traditional $\forall$ and $\exists$ Predicate Logic quantifiers, which is why we have the $\sf NQ$ class. For example, imagine a program ${\sf cntSbyS}(s)$ that takes a string $s$ containing properly nested and possibly side-by-side brackets as input, and count how many side-by-side groups $s$ contains. The pre-condition essentially requires $s$ to satisfy the Context Free Grammar:
\begin{equation}\label{ex:prodrule}
S \ \rightarrow \
         "" \ \ | \  \ "(" \; S \; ")" \; S
\end{equation}
While simple, this is not expressible in Predicate Logic due to the lack of recursion. To express this pre-condition in Python we can for example write this:

{\small
\[
\begin{array}{l}
{\color{violet}\bf def}\; {\sf preCntSbyS}(s): \; {\color{cyan}\bf return}\ {\sf S}(s) == "" \\ 
{\color{violet}\bf def}\; {\sf S}(s): \  \mbox{\color{pink}\# for recognizing $S$} \\
\hspace{3mm}   
    \begin{array}[t]{llcl}
    {\bf if} & s==""\; {\bf or}\; s=={\sf None} & : &{\color{cyan}\bf return} \; s \\
%\hspace{3mm}  
    {\bf else} & & : & 
    {\color{cyan}\bf return}\ {\sf S}({\sf close}({\sf S}({\sf open}(s))))
    \end{array}\\
{\color{violet}\bf def}\ {\sf open}(s): 
\ \mbox{\color{pink}\# for recognizing "("} \\
\hspace{3mm} 
    \begin{array}[t]{llcl}{\bf if} & s==""\; {\bf or}\; s=={\sf None} & : & {\color{cyan}\bf return} \; {\sf None} \\
    {\bf elif} & s[0]=="(" & : &{\color{cyan}\bf return} \  s[1{:}] \\
    {\bf else} & & : &  {\color{cyan}\bf return} \ {\sf None}    \end{array}\\
{\color{violet}\sf def}\; {\sf close}(s) : \   \mbox{\color{pink}\# analogous to $\sf open$, for recognizing ")"} \\
\ 
\end{array}
\]
}  

Notice that the recursion in the code of $\sf S$ reflects the recursion in the grammar rule in (\ref{ex:prodrule}).
The above post-condition is of the difficulty class $\sf NQ$. %($di\!f = 8$).

We can also look at the 'structural' difficulty of a pre and post-condition from the perspective of the code complexity of functions that formulate them, expressed as  McCabe (cyclomatic) complexity  (CC)~\citep{mccabe1976complexity}. For example the above pre-condition for $\sf cntSbyS$ has a CC of 11. So, in addition to being semantically difficult (classs $\sf NQ$), the pre-condition is also quite complex from the code complexity perspective.

\section{Benchmarking Methodology} \label{sec:methodology}
% \todo[inline]{Could we consider to explicitly call the section Benchmarking Methodology? \XCOMMENT{I think that would be a more proper naming}}

Here we describe our approach for benchmarking LLMs with respect to the task of generating valid pre- and post-conditions. {\em For simplicity  we talk about post-conditions}, but the same considerations apply to pre-conditions. Peculiarities in  validation steps are explained later in Section~\ref{sec:validation}.

\subsection{Prompting}

% Once we pre-processed the data, we could start generating post-conditions using different LLM. Here, two approaches could be considered.

% \begin{itemize}
%     \item \emph{Plain} Give the prompt to the LLM as it is. Repeat $n$ times for statistical robustness. 
%     \item \emph{Evolving} Use LLM in an interactive way. That is
%     \begin{enumerate}
%         \item prompt \textsf{P} to the LLM that return post-condition $\mathsf{PC}_1$
%         \item run tests (manual? generated?) on $\mathsf{PC}_1$
%         \item if all tests pass then conclude
%         \item if some test fails, create a new prompt \textsf{P'} that signal the error (to evaluate how) and give it to the LLM
%         \item iterate until all tests pass or the maximum number of iterations is reached
%     \end{enumerate}
%     When evolution/iteration ends, use validationSuite to evaluate the obtained post-condition.
% \end{itemize}

To obtain a formal post-condition from an LLM, we prompt the LLM as shown in Figure \ref{fig:prompt}. 
%the prompt for pre-conditions is similar.
\NEWB{The wordings are formulated by considering 
how developers could reasonably be expected to formulate them, and what information could reasonably be included in the prompt. To avoid bias, we do not engineer the prompt towards certain LLMs. The program code is not included in the prompt as in this study we focus on inference from the program's description.}

The prompt contains a natural language description of the program and few examples of input and output. This is  followed by instructions:

%The prompt {\em does not} include the program's code. This is deliberate as in this study we want to focus on formalizing specifications from natural language descriptions; more on this will be discussed in the discussion Section \ref{sec:discussion}. 
%Then, it gives the following instruction.
\begin{enumerate}[leftmargin=*]
    \item The LLM is first asked to extract the post-condition of the program in English. The rationale is that a program's description, as exemplified by the description of the program $sortx$ shown in Section \ref{sec.intro}, describes the program; but in most cases it is {\em not}  structured into sections explicitly called pre- and post-conditions. So we ask the LLM to first explicitly formulate the post-condition.
    \item Then, the second instruction asks the LLM to code the extracted post-condition as a Python (boolean) function of a specific signature.
\end{enumerate}
Notice that we instruct the LLM to express the post-condition as a {\em  function}, different from other work that instructs the LLM to write assertions~\citep{endres2024can,he2024beyond}, for example: 
\begin{quote}
"{\em Write a symbolic postcondition for $P$ consisting of exactly one {\bf assert statement}}" \citep{endres2024can}.
\end{quote}
This might be constraining the LLM. For human programmers, it is not always convenient, or even possible, to code a {\em full} post-condition in a single $\sf assert$.
% This depends on how complex the post-condition is. 
Since LLMs are trained from human data, this may influence the resulting LLMs' behavior.
In contrast, we phrase the problem as a task of writing a function, with the rationale that this would relate better to LLMs' general programming training/skill.

\begin{figure}
\centering
\fbox{\small
\begin{minipage}{\columnwidth - 5mm}
Consider a program \colorbox{yellow}{\em P}. \\ 
\colorbox{pink}
{\em Description of what the program does}. \\

Examples: \\
  $P$({\em inputs}) = {\em ouput} \\
  $P$({\em inputs}) = {\em ouput} \\
  ... \\

INSTRUCTION:\\
(1) {\color{blue}\bf 
 Extract} the {\color{red}post}-condition of $P$ (in English). \\
(2) Then, {\color{blue}\bf code} this {\color{red}post}-condition {\color{blue} as a Python function} with the header shown below, where the parameter r represents $P$'s return value.\\
Do not explain. If a helper function is needed, define it as an inner function. Import packages, if needed, locally within the function. \\

{\footnotesize
def check\_post\_$P$($r{:}type, par_1{:}type, par_2{:}type,...$) -> bool {:}}\\
\end{minipage}
}   
\vspace{2mm}
\caption{\em The prompt to get a post-condition from an LLM.
$P$ is to be replaced with the name of the program whose post-condition we want to extract. The \colorbox{pink}{red-part} is to be filled with a textual, natural language, description of $P$'s functionality. Several examples of inputs-outputs of $P$ are also included in the prompt, as part of $P$'s description.
%To get a pre-condition, a new prompt is sent, with the same text as above, but with the word 'post' replaced by 'pre'. 
}
\label{fig:prompt}
\end{figure}

To obtain multiple answers from the LLM, we send the prompt multiple times, and collect the answer of each.
\NEW{The same prompt is used across all subject LLMs (Section \ref{sec:models})\footnote{Tuning the prompt per model would introduce confounding variables related to optimization resulting in a wide configuration space which goes well beyond the scope of in this paper.}
}.
Despite the instruction that says: "{\em Do not explain}", an LLM can still insist to be helpful by giving explaining text along with the code. 
While such an answer is not what the prompt asks, a pre-processing step is still applied to try to locate the code in the text and extract it. It may also happen that the generated Python code is not properly indented (e.g. without any indentation at all), making the code syntactically incorrect. The pre-processor also makes an attempt to fix the indentation. This pre-processing is uniformly applied to every answer from any LLM.

\subsection{Solution validation} \label{sec:validation} %\todo{this subsection could be a bit difficult to follow, we could consider using a sort of flowchart diagram?}

Suppose for a program $F(x)$ an LLM proposes a pre-condition $pre'$ and a post-condition $post'$. \NEW{These proposals are {\em correct if they are semantically equivalent} to the provided {\bf solution} $pre_F$ and $post_F$.}
This is validated through testing. 
\NEW{Validation through formal verification is not a realistic option as many of the conditions are complex and fall beyond the range of automated theorem provers, as in the case of the example program $\sf cntSbyS$ shown in Section \ref{sec.spec.diff}.
Also, note that we do {\bf not} check for mere syntactical similarities, as there are infinitely many ways the same condition can formulated.} %, and still semantically means the same thing.}

Test suites $T_{pre}, T_{post}$ are provided by the dataset (Section~\ref{sec:dataset}). Each test case $x \in T_{pre}$ represents an input for $F$.  
Each test case in $T_{post}$ is a pair $(o,x)$ where $x$ represents an input for $F$ and $o$ a value that could be $F$'s output.

\begin{itemize}[leftmargin=*]
    \item The LLM proposed pre-condition $pre'$ is accepted as a correct pre-condition for $F$ if for all $x\in T_{pre}$, $pre'(x) = pre_F(x)$.
    \item The LLM proposed post-condition $post'$ is accepted as a correct post-condition for $F$ if for all $(o,x)\in T_{post}$, $post'(o,x) = post_F(o,x)$.
\end{itemize}
 
Note that $x$ can represent an invalid input for $F$, for which $pre_F(x)$ will give a $f\!alse$ to reject it. Similarly, for a valid input $x$, $o$ could represent a wrong output by $F$ for which $post_{F}(o,x)$ will give a $f\!alse$ to reject this $o$. Such tests are {\bf negative} tests. It is important that the test suites in the study include negative tests (they do) to avoid accepting proposals that are too weak.
The test suites in the dataset were manually written and achieve full branch coverage over their corresponding programs. Still, to increase the variety of the tests and to maximize the likelihood of detecting corner cases that challenge LLMs, we developed an automated test generation pipeline that uses Pynguin~\citep{Pynguin2020}, the Python General Unit test generator, and Poodle~\citep{poodle}, a tool for mutation testing of Python code.

The basic workflow without Poodle in shown in Figure \ref{fig:flowchart1} ---we will go back to Poodle later.
Given a program  $F(x)$, we first run Pynguin on $F(x)$, to obtain test cases $t^1_F \dots t^n_F$, \NEW{each is an instance of $x$.} Then:

\begin{figure*}
    \centering
    \includegraphics[scale=0.3]{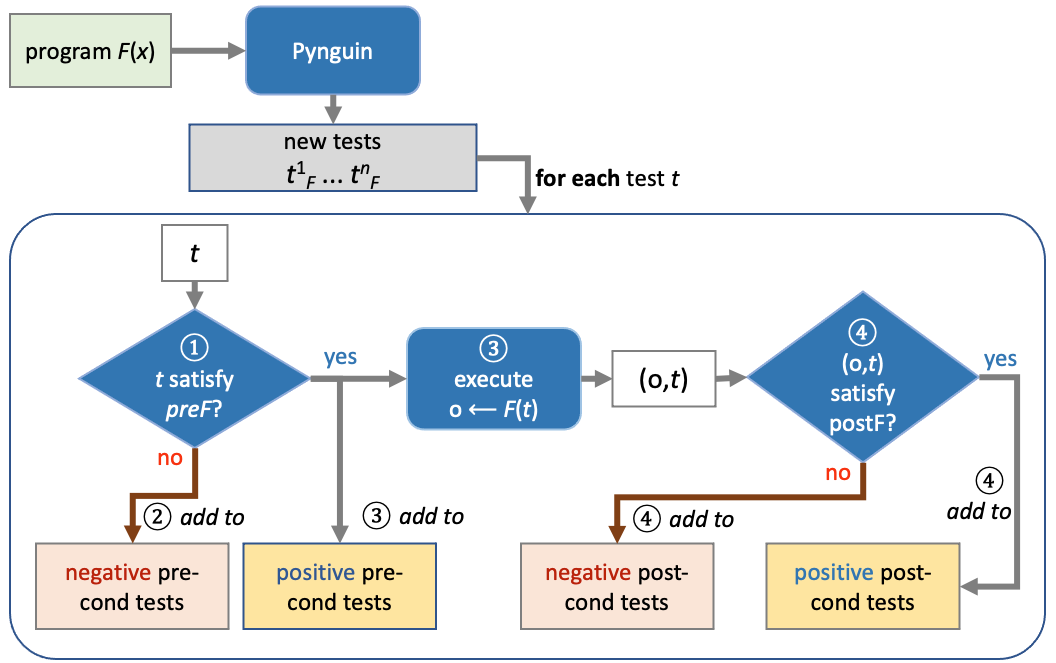}
    \caption{\em The basic workflow for generating more positive and negative tests using Pynguin. $F(x)$ is the program under consideration; $preF$ and $postF$ are solution pre- and post-conditions. Using $F$ we cannot actually generate negative post-condition tests. To do the latter, $F$ is replaced by a mutant ($F$ containing a mutated code) and re-run the same workflow.}
    \label{fig:flowchart1}
\end{figure*}

\begin{enumerate}[leftmargin=*]
    \item For each test case $t^i_F$, we evaluate the solution pre-condition on the test case:  $pre_F(t^i_F)$.
    
    \item If $pre_F(t^i_F)$ evaluates to $f\!alse$, we add $t^i_F$ to the set of {\em negative} pre-condition tests for $F(x)$. 

    \item Otherwise, if $pre_F(t^i_F)$ evaluates to $true$,
    we classify $t^i_F$ as a {\em positive} pre-condition test. 
    Then, we also execute the program $F(x)$  using the input $t^i_F$ to obtain the output $o^i_F$. 

    \item We then evaluate the solution post-condition: $post_F(o^i_F, t^i_F)$ and add the pair $(o^i_F, t^i_F )$ to the post-condition test suite of $F(x)$.

\end{enumerate}

\NEW{
Because in step-3 above we use $F(x)$ itself to obtain $o^i_F$, when in step-4 we evaluate $(o^i_F,t^i_F)$ using $post_F$, it will always result in $true$. So, tests added in step-4 are always {\em positive} post-condition tests, and therefore with the above scheme Pynguin cannot add negative post-condition tests. 
To be able to do the latter, we generate mutated versions $F_1(x), \ldots, F_k(x)$ of the program $F(x)$ using Poodle. 
For each mutant $F_j(x)$, we execute Pynguin again and repeat the
workflow in Figure~\ref{fig:flowchart1} to categorize the generated test cases into the appropriate test suites:  {\em positive} and  {\em negative} for both pre- and post-conditions. 
In particular, if in step-4 evaluating $post_F(o^i,t^i_F)$ gives $false$ (where $o_i$ is produced by a mutant of $F$), the test $(o^i,t^i_F)$ is thus a {\em negative} post-condition test.
}
 
% Since Pynguin is designed to cover nominal paths  of the input program $F(x)$, it cannot, by design, generate {\em negative} tests for the post-condition. To explore alternative paths, we generate mutated versions $F_1(x), \ldots, F_k(x)$ of the program $F(x)$ using Poodle. 
% %
% \XCOMMENT{Wish: what is a 'nominal' path? I actually don't know this terminology. I think it is more due to the fact that we use $F$, which is assumed to be correct, to produce $o^i_F$, that is responsible for not being able to produce negative post-cond test. Proposing the following reformulation:

% "Because in step-3 above we use $F(x)$ itself to obtain $o^i_F$, when in step-4 we evaluate $(s^i_F,t^i_F)$ using $post_F$, it will always result in $true$. So, tests added in step-4 are always {\em positive} post-condition tests, and therefore with the above scheme Pynguin cannot add negative post-condition tests. 
% To able able to do the latter, we generate mutated versions $F_1(x), \ldots, F_k(x)$ of the program $F(x)$ using Poodle. "
% }
% %
% For each mutant $F_j(x)$, we execute Pynguin again and repeat the previously described procedure to categorize the generated test cases into the appropriate test suites:  {\em positive} and  {\em negative} for both pre- and post-conditions. 
% \NEW{In particular, if in step-4 evaluating $post_F(o^i,t^i_F)$ gives $false$ (where $o_i$ is produced by a mutant of $F$), the test $(o^i,t^i_F)$ is thus a {\em negative} post-condition test.}

%Clarify differences between pre and post tests

%\subsubsection{Test suite extension}

\section{Experiment Design} \label{sec:experiment-design}

This section presents the design of the empirical study we performed to answer the research questions, which we describe first, followed by the dataset designed for the study, as well as the different models considered. Finally, we will describe the experiment protocol and  evaluation metrics. 
%The results of the experimental study will be discussed next in Section~\ref{sec:results}.

\subsection{Research questions}
The principal research questions that drive the empirical investigation concern the performance of LLMs in producing correct \emph{full} pre- and post-conditions from descriptions in natural language (RQ1), and the impact of the difficulty of the task on the performance of the LLMs (RQ2). Specifically:

\begin{enumerate}[leftmargin=*]
    \item {\bf RQ1}: {\em how do LLMs perform in generating formal full pre- and post-conditions? How good are open source LLMs compared to proprietary ones?} 
    %We further refine this research question into two sub-questions in which we investigate the performance of the LLMs with respect to the generation of pre-conditions ({\bf RQ1.1}) and and post-conditions ({\bf RQ1.2}).
    \item {\bf RQ2}: {\em how do LLMs perform on pre-/post-conditions of varying levels of difficulty? }
    % For example, a post-condition that requires a quantifier to express is more challenging to write for a human programmer than one that does not need a quantifier. A post-condition that requires nested quantifiers would be even more challenging to write. There are also programs whose post-condition is hard, or even impossible, to express with basic quantifiers a la Predicate Logic.
    % We investigate whether LLMs can generate pre-/post-conditions with varying levels of difficulty. 
    % If they can, the skill would be very helpful to aid programmers. 
    
    We refine RQ2 into  following sub-questions based on the two notions of difficulty discussed in Section~\ref{sec:prepost}: \\
    {\bf RQ2.1} cyclomatic complexity: how much does the cyclomatic complexity of the pre-/post-condition impact the performance of LLMs?  %For this RQ, we compute the cyclomatic complexity of the pre-/post-condition code in our dataset and grouped them into two: those with low complexity (cyclomatic complexity less than or equal to 5) and those with high complexity (cyclomatic complexity above 5).
    {\bf RQ2.2} quantifiability: how much does the semantic difficulty level ($\sf S$, $\sf Q$, $\sf QQ$, $\sf NQ$) impact LLMs' capability in formalizing contracts? 
        %\todo[inline]{describe the four levels here \XCOMMENT{this was in SIII}}
\end{enumerate}

\subsection{Dataset} \label{sec:dataset}

\NEWB{Previous studies~\citep{endres2024can,he2024beyond} use the HumanEval dataset~\citep{chen2021evaluating}. This dataset is publicly available, so it might have been included in the training of the studied LLMs, which would then bias the results.
To mitigate this, our dataset,} called {\sf HEx}, consists of 40 tasks {\bf freshly adapted} from the HumanEval dataset~\citep{chen2021evaluating}. The repository of 
{\sf HEx} is \NEWB{unexposed} and unlikely to have been used for the training of an LLM.
\NEWB{The 40 tasks were selected with consideration to provide a balanced distribution of difficulty levels for pre- and post-conditions. Table \ref{tab:HEx} shows the distribution, reflecting a well-spread range of difficulty levels. 
Importantly,  no consideration was made towards the strength or weakness of any particular LLM, so as not to bias the selection towards a particular LLM.}

Each task ${\sf HEx}_k$ is based on the task ${\sf HE}_k$ in HumanEval. The new task either paraphrases the original natural language description and adds some new nuances/tweaks, or it formulates a related task, rather than an extension of the original ${\sf HE}_k$. 
Note that for LLMs, paraphrasing matters, as they are trained on natural language input~\citep{guan2025order}.
The idea is that even if an LLM has been trained with HumanEval, it needs to be able to generalize to solve $\sf HEx$ tasks.
For example, below is the original description of ${\sf HE}_{104}$ and its modified ${\sf HEx}_{104}$. Notice that the phrasing is different. The added twist is marked in bold. 

\begin{quote}
\hspace{-7mm}
\begin{tabular}{|l|m{60mm}|} \hline
${\sf HE}_{104}$ &
   {\em Given a list of positive integers $x$, return a sorted list of all    elements that hasn't any even digit. Note: Returned list should be sorted in increasing order.} \\ \hline
${\sf HEx}_{104}$ & 
   {\em Given a non-empty array $s$ of integers and an integer $k$ between 1 and 9 (inclusive), the program returns a sorted array $t$ consisting of all integers in s {\bf which have a digit that is divisible by $k$}. Digit 0 is considered as always divisible.} \\ \hline
\end{tabular} 
\end{quote}
 
\vspace{3mm}
\noindent
Each {\sf HEx} task is a tuple:
\begin{equation}
(D,\;F,\;pre_F,\;post_F,\;T_{pre},\;T_{post})  
\end{equation}
\begin{itemize}[leftmargin=*]
    \item $D$ is a natural language specification of some program,
    \item $F$ is a Python implementation of $D$,
    \item $post_F$ is a Boolean function (also in Python) formalizing $F$'s {\bf full} post-condition,
    \item $pre_F$ is a Boolean function formalizing $F$'s {\bf full} pre-condition,
    \item $T_{pre}$ and $T_{post}$ are test suites. The first is used to validate proposed (by LLMs) pre-conditions, the second to validate proposed post-conditions.
\end{itemize}

Given $D$, the task for an LLM is to come up with $pre'$ and $post'$ which are equivalent to the provided solution $pre_F$ and $post_F$. Since full verification of equivalence is undecidable, the test suites $T_{pre}$ and $T_{post}$ are used instead to validate this (see Section~\ref{sec:validation}). The original HumanEval was meant to assess LLMs' ability to construct programs given their natural language description ($D$); so, $pre_F$ and $post_F$ solutions were not provided. In our dataset $\sf HEx$, these components are provided.

All $\sf HEx$ tasks have post-conditions; 19 have pre-conditions (the rest have no constraint on their inputs). We take the difficulty of their solutions,  $pre_F$ and $post_F$, as an indicator of their actual difficulty.
Table \ref{tab:HEx} shows the distribution of the difficulty class (see Section~\ref{sec:prepost}) and complexity (of the solutions). 
%The column $di\!f$ shows the average difficulty number;  
%The column CC shows the average complexity number.
The difficulty class is manually classified by a human expert. The cyclomatic complexity is calculated with the metric tool 
%Radon 
\cite{radon}.
%the code metric tool Radon.
The pre-conditions are generally simple, though there are two $\sf NQ$s. 
For example, the task ${\sf HEx}_1$ describes a program that takes a string input satisfying the grammar in (\ref{ex:prodrule}). As previously discussed, such a pre-condition is $\sf NQ$.
Almost 75\% of the post-conditions are challenging ($\sf QQ$ or $\sf NQ$).
For example, the solution post-condition of the aforementioned ${\sf HEx}_{104}$ is of the difficulty class $\sf QQ$. 
%

% $T$ consists of manual tests, with 100\% branch coverage over $F$. To reduce bias, $T$ is also filled with tests generated by the tool Pynguin \cite{Pynguin2020}. Over the whole dataset, there are 162 human tests for the pre-conditions, and 603 for the post-conditions. These are augmented with 1383 Pynguin tests for pre-conditions, and 1662 for post-conditions. So, in average there are 64 tests per condition. 
% In average, 14 of them are negative tests, representing pre-condition checks against invalid inputs and post-condition checks against wrong $F$'s outputs; AI-written $pre'$ and $post'$ should thus return {\em false}.  

\begin{table}
\caption{Distribution of the difficulty and complexity of $\sf HEx$ pre- and post-conditions.}\label{tab:HEx}
\centering
{\footnotesize
\begin{tabular}{|c|c|c|c|c|c|c|c|c|} \hline
  & & \multicolumn{4}{c}{difficulty} 
  & \multicolumn{3}{|c|}{cyclomatic complexity} \\ \cline{3-9}
  & $N$ & $\bf S$ & $\bf Q$ & $\bf QQ$ & $\bf NQ$ & 
  %$di\!f$ & 
  low & high & avg. CC \\ \hline
pre-c & 19 & 10 & 7 & 0 & 2 & 
%2.1 & 
17 & 2  & 2.8 \\ \hline
post-c & 40 & 5 & 7 & 11 & 17 & 
%5.0 & 
25 & 15  & 5.4 \\ \hline 
\end{tabular}
}
\end{table}

%\todo[inline]{Consider to move  Prompting section here, before models description and to remove methodology.}

\subsection{Models}\label{sec:models}

The EvalPlus Leaderboard~\citep{EvalPlusLeaderboard} provides a comparison of various LLMs designed for code generation, ranking them based on the pass@\!1 metric \citep{chen2021evaluating}. For our selection process, we focused on models with a pass@\!1 score greater than 70.
We downloaded open models from Hugging Face~\citep{huggignface} and eliminated those that could not be processed using llama.cpp~\citep{llamacpp} or that exceeded our computational resources. The final list of open models, shown in light blue in Table~\ref{tab:models}, also includes some recent releases such as Athene and Qwen3, resulting in 17 models.
The closed models primarily come from OpenAI\textsuperscript{\textcopyright}, Google\textsuperscript{\textcopyright}, and Anthropic\textsuperscript{\textcopyright}, totaling 7 models (represented in dark blue in Table~\ref{tab:models}). This selection features at least two generations from each model family. Overall, Table~\ref{tab:models} presents a complete list of 24 models used in our study, detailing their full names, abbreviated names used in plots, and, for the open models, the number of parameters and size.

\begin{table*}[htb]
\caption{LLMs used in the study. Proprietary models are shaded in blue, open models in light blue. }
\label{tab:models}
\centering
\begingroup
%\fontsize{5pt}{6pt}\selectfont
\begin{tabular}{llrr}
  \hline
Full name & Short name & Parameters (B) & Size (GB) \\ 
  \hline
   \cellcolor[HTML]{3690c0} claude-3-7-sonnet & Claude3.7 & - & - \\ 
  \cellcolor[HTML]{3690c0} claude3-haiku & Claude3 & - & - \\ 
  \cellcolor[HTML]{3690c0} gemini-1.5-flash & Gemini1.5 & - & - \\ 
  \cellcolor[HTML]{3690c0} gemini-2.0-flash-exp & Gemini2.0 & - & - \\ 
  \cellcolor[HTML]{3690c0} gpt-3.5-turbo & GTP3.5 & - & - \\ 
  \cellcolor[HTML]{3690c0} gpt-4-turbo & GPT4 & - & - \\ 
  \cellcolor[HTML]{3690c0} gpt-4o & GPT4o & - & - \\ 
  \cellcolor[HTML]{a6bddb} athene-v2-chat & Athene & 72.70 & 271 \\ 
  \cellcolor[HTML]{a6bddb} CodeFuse-DeepSeek-33B & CodeFuse & 33.30 & 125 \\ 
  \cellcolor[HTML]{a6bddb} codegemma-7b-it & CodeGemma & 8.54 &  32 \\ 
  \cellcolor[HTML]{a6bddb} CodeQwen1.5-7B-Chat & CodeQwen1.5 & 7.25 &  28 \\ 
  \cellcolor[HTML]{a6bddb} Codestral-22B-v0.1 & Codestral & 22.20 &  83 \\ 
  \cellcolor[HTML]{a6bddb} deepseek-coder-33b-instruct & DeepSeekCoder & 33.30 & 125 \\ 
  \cellcolor[HTML]{a6bddb} gemma-3-27b & Gemma3 & 27.40 & 101 \\ 
  \cellcolor[HTML]{a6bddb} Hermes-2-Theta-Llama-3-70B & Hermes2 & 70.60 & 263 \\ 
  \cellcolor[HTML]{a6bddb} Meta-Llama-3-70B-Instruct & Llama3 & 70.60 & 263 \\ 
  \cellcolor[HTML]{a6bddb} Nxcode-CQ-7B-orpo & Nxcode & 7.25 &  28 \\ 
  \cellcolor[HTML]{a6bddb} OpenCodeInterpreter-DS-6.7B & OpenCodeInt & 6.74 &  26 \\ 
  \cellcolor[HTML]{a6bddb} OpenCoder-8B-Instruct & OpenCoder & 7.77 &  29 \\ 
  \cellcolor[HTML]{a6bddb} Qwen2.5-Coder-32B-Instruct & Qwen2.5 & 32.80 & 123 \\ 
  \cellcolor[HTML]{a6bddb} speechless-codellama-34b-v2.0 & Codellama & 34.00 & 126 \\ 
  \cellcolor[HTML]{a6bddb} starchat2-15b-v0.1 & Starchat2 & 16.00 &  60 \\ 
  \cellcolor[HTML]{a6bddb} WizardCoder-33B-V1.1 & WizardCoder & 33.30 & 125 \\ 
  \cellcolor[HTML]{a6bddb} Qwen3-32B & Qwen3 & 32.80 & 123 \\ 
   \hline
\end{tabular}
\endgroup
\end{table*}

% \HIDE{
% \begin{table}
% \caption{Models}
% \centering
% {\scriptsize
% \begin{tabular}{|c|c|c|} \hline
%   \cellcolor[HTML]{11aadd} GPT3.5 & 
%   \cellcolor[HTML]{11aadd} GPT4 & 
%   \cellcolor[HTML]{11aadd} GPT4o  \\ \hline
%   \cellcolor[HTML]{11aadd} Claude3 &
%   \cellcolor[HTML]{11aadd} Claude3.7 & 
%   \cellcolor[HTML]{11aadd} Gemini1.5 \\ \hline
%   \cellcolor[HTML]{11aadd} Gemini2.0 & 
%   Athene2 (??) & 
%   Gemma3 (27B) \\ \hline
%   Codestral (22B) &
%   Llama3 (70B) &
%   CodeQwen1.5 (7B) \\ \hline 
%   Qwen2.5-Coder (32B)  
% \end{tabular}
% }
% \end{table}
% }

\subsection{Experiment protocol} \label{sec:exper:protocol}

The experimental protocol follows the methodology described in Section~\ref{sec:methodology} and comprises two parts: prompting and solution validation. 

During the prompting phase, we send the prompt 10 times for each model and each task, resulting in 4,560 proposed solutions for pre-conditions and 9,600 proposed solutions for post-conditions.
Proprietary models are accessed through their APIs, while open models are run locally on a machine equipped with 8 Tesla\textsuperscript{\textcopyright} A40 GPUs, each with 48GB of VRAM. The size of open models is limited to 380GB, which is the total amount of VRAM available when using all 8 GPUs simultaneously. All models operate with a temperature setting, controlling the randomness of the model's output, of 0.7.
\NEWB{This is the typical default of many LLMs, giving a medium level of randomness.} The maximum number of tokens, which limits the length of the generated text, is set to 1,024. \NEWB{Recall that the prompt specifically asks the LLM to just answer with the code of pre-/post-conditions; the limit should be generous for this.}

The solution validation phase comprises of 162 manually written tests for pre-conditions and 603 for post-conditions.
For each program $F$ in the dataset, the manually written tests  achieves 100\% branch coverage. To provide additional strength, 
these test suites are extended further using the tools Pynguin and Poodle, as detailed in Section~\ref{sec:validation}.
Pynguin is configured with a maximum test generation time of 180 seconds and uses the DYNAMOSA algorithm.
As all programs in the dataset use Python type annotations, we emphasize explicit typing by setting {\em original_\!type_\!weight} to 100, while minimizing the weights of inference parameters {\em type4py} and {\em type_\!tracing} to 0.01. The tool targets {\em branch} coverage and limits {\em collection size} to 10 elements. Finally, a low {\em object_\!reuse_\!probability} of 0.01 promotes the creation of new objects, increasing test diversity. In total, we generated additional 1383 tests for pre-conditions and 1662 for post-conditions. 

% Pynguin parameters
%             "--maximum_search_time", 180,
%            "--original_type_weight", "100.0",
%            "--type4py_weight", "0.01",
%            "--type_tracing_weight", "0.01",
%            "--any_weight", "0.01",
%            "--collection_size", "10",
%            "--none_weight", "0.01",
%            "--object_reuse_probability", "0.01",
%            "--algorithm", DYNAMOSA,
%            "--coverage-metrics", "BRANCH",

% Poodle parameters: dafult
   
% $T$ consists of manual tests, with 100\% branch coverage over $F$. To reduce bias, $T$ is also filled with tests generated by the tool Pynguin \cite{Pynguin2020}. Over the whole dataset, there are 162 human tests for the pre-conditions, and 603 for the post-conditions. These are augmented with 1383 Pynguin tests for pre-conditions, and 1662 for post-conditions. So, in average there are 64 tests per condition. 
% In average, 14 of them are negative tests, representing pre-condition checks against invalid inputs and post-condition checks against wrong $F$'s outputs; AI-written $pre'$ and $post'$ should thus return {\em false}.  

%including Qwen, Deepseek Coder,, Google's Gemini, and Anthropic's Claude

\subsection{Metrics} \label{sec:metrics}

% This section outlines the metrics we compute to address our research questions, based on the results derived from the protocol described earlier.

To evaluate LLM performance (\textbf{RQ1}), we measure different aspects of the quality of the solution proposed by the LLM. 
\NEW{In total, seven metrics are used. In previous work \citep{wen2024enchanting,endres2024can,he2024beyond} much emphasis was put on the so-called accept@\!1 metric, similar to the popular pass@\!1 \citep{chen2021evaluating}. We propose more refined metrics that also give further insight on how close are LLMs' solutions to being good, allowing for a more nuanced understanding of the strengths and weaknesses of LLMs in formalizing pre- and post-conditions. 
}
\\

\noindent
{\bf \acceptone}\\
%
% For a given task and an LLM, accept@\!1~\citep{endres2024can} is defined as the expected probability of extracting a correct solution when sampling one solution from a set of 10 replicas. Accept@\!1 corresponds to the proportion of correct solutions observed in 10 replicas. 
% \XCOMMENT{WP: I propose to reformulate it a bit, to separate the metric itself from how many replicas we have. How about something like this, and also to give a bit more context:
%
\NEW{For a given task and an LLM, the metric accept@\!k is conceptually defined as the expected probability of extracting a correct solution when sampling $k$ solutions \citep{endres2024can}. In particular, accept@\!1 is then the probability of getting the right answer from just one attempt. In our experiment setup, see previous Section \ref{sec:exper:protocol}, for every task the prompt is repeated ten times. So, we get ten 'replicas'. The expected probability accept@\!1 is approximated by the proportion of correct solutions observed among these ten replicas.
}
\\

\noindent
{\bf Mean Passing Tests} % per task}

\NEW{When an LLM fails to give a correct solution, near-correct solutions could still be valuable. For instance a developer could accept a near-correct solution and make an effort to fix it. Hence, we also want to know how close the solutions proposed by the LLM are to being correct. The accept@\!1 metric is however too coarse for this, as it only looks at correct solutions.}
%
%Accept@\!1 provides a coarse differentiation of models, as it does not account for the proportion of passing tests.
%
Suppose that for a program $F(x)$ models $M_1$ and $M_2$ return solutions $s_1$ and $s_2$. If $s_1$ passes 99\% of the tests and $s_2$  5\%, we could say that $M_1$ generates a better approximation of the correct solution than $M_2$. Consequently we introduce the \emph{Mean Passing Tests} metric which captures this notion. For a given LLM and a task $T$, \emph{Mean Passing Tests} is computed as the average proportion of correct (passing) tests across 10 replicas for the task $T$. \\

\noindent
{\bf \errorrate}

With the \errorrate metric we quantify  
%\todo{I think we should focus on the outcome of the pre/post condition rather than on the test, maybe avoid altogether naming tests?} 
all scenarios where an LLM proposed pre- or post-condition $p$ fails to produce a \NEW{well-formed} output. 
%\XCOMMENT{WP: I propose to use another term than "valid" as the word is somewhat overused, I think. How about "well formed output"? DP: agree. I'll update text and figures} 
Examples of this include cases where $p$ is not present at all, or $p$ is present but it does not return a boolean value when applied to a test $t$ or where $p$ is not even syntactically correct, and will thus crash when applied to $t$.  
Unlike failed tests, which require an oracle for detection, an \emph{error} could potentially be identified by examining the output of the LLM proposed solution. For a given LLM, the \errorrate metric is computed as the number of runs (out of 10) that resulted in error. 
\NEW{For an LLM, a high \errorrate is an indication that it struggles in the more basic skill of producing syntactically and type correct pre-/post-conditions.} \\

%\todo[inline]{link to the definition of positive and negative tests}
\noindent
{\bf False positive and false negative rates} 
%\XCOMMENT{WP: maybe we should discuss again whether or not to reverse the polarity :) I would argue that in software testing/verification, a false positive usually means a test  reports an error, while it is actually not an error.}

If a test does not result in an \emph{error}, the pass/fail classification can be  further refined by considering whether it is a \emph{positive} or a \emph{negative} test, as defined in Section~\ref{sec:validation}.
Consider a program $F$ that takes an input $x$ and produces an output $F(x)$. A post-condition is a boolean function, denoted as $post_F(F(x),x)$, that formally specifies the properties the output $F(x)$ must satisfy for a given input $x$ for the program $F$ to be considered correct. A test case for the post-condition  $post_F$ is a pair $(o,x)$, where $x$ is a valid input for the program $F$ and $o$ is a possible output of $F$. The test $(o,x)$ is a \emph{positive} test if $post_F(o,x)$ evaluates to \emph{true}, else $(o,x)$ is a \emph{negative} test. 

As an example, let us consider the program {\sf findMax} intended to find the maximum value in a non-empty list {\sf a} of integers. 
A robust post-condition {\sf post}  for {\sf findMax} would assert the following two properties: the returned value must be present in the input list, and the returned value must be greater than or equal to every other element in the list (see Eq.~\ref{spec:findMax}).
A Python function that encodes a correct post-condition is shown below:

{\small
\[
\begin{array}{ll}
{\color{violet}\bf def} & {\sf post}(retval{:}{\color{violet} int}, \; a{:}{\color{violet} list}[{\color{violet} int}]) \rightarrow  {\color{violet} bool} : \\
   &
   {\color{cyan}\bf return}\  retval\;{\sf in}\; a
   \ {\sf and}\ {\sf all}(retval \geq x \;{\bf for} \;x\; {\bf in}\; a)
\end{array}
\] }

\HIDE{
{\scriptsize
\begin{lstlisting}[language=Python]
def post(retval:int, a:list[int]) -> bool:
  if not a:
    return False
  return retval in a and all(retval >= val for val in a)
\end{lstlisting}}
}

\noindent The pair $(-2, [-5, -2, -10])$ is a \emph{positive} test for {\sf post}, because a correctly implemented {\sf findMax} should return $-2$ for the input list $[-5, -2, -10]$. 
Now suppose that the model  $M_1$ incorrectly assumes that the output must be non-negative and returns the following buggy post-condition:

{\small
\[
\begin{array}{ll}
{\color{violet}\bf def} & {\sf post\_\!M1}(retval{:}{\color{violet} int}, \; a{:}{\color{violet} list}[{\color{violet} int}]) \rightarrow  {\color{violet} bool} : \\
   &
   {\color{cyan}\bf return}\  retval\;{\sf in}\; a
   \\ 
   & \hspace{10mm} {\sf and}\ {\sf all}(retval \geq x \;{\bf for} \;x\; {\bf in}\; a) \ {\sf and}\  retval \geq 0
\end{array}
\] }

\HIDE{
{\scriptsize
\begin{lstlisting}[language=Python]
def post_M1(retval:int, a:list[int]) -> bool:
  if not a:
    return False  
  return retval in a and all(retval >= val for val in a) and retval >= 0
\end{lstlisting}}
}

\noindent The post-condition {\sf post\_\!M1} evaluates to \emph{false} on the \emph{positive} test  $(-2, [-5, -2, -10] )$. We call this a {\bf false negative}, as {\sf post\_\!M1} incorrectly states that $-2$ is not the correct output for the input $[-5, -2, -10]$ and thus marking a correct implementation of the {\sf findMax} program as incorrect.

Consider now a \emph{negative} test for {\sf post}: $(20, [5, 2, 10])$. This test is negative because  an implementation of {\sf findMax}  that returns $20$ for the input list $[5, 2, 10]$ is incorrect. Therefore, the correct post-condition {\tt post} evaluates to \emph{false}.
Consider a different model, $M_2$, which does not recognise that the output should be an element of the input list and produced the following post-condition:

{\small
\[
\begin{array}{ll}
{\color{violet}\bf def} & {\sf post\_\!M2}(retval{:}{\color{violet} int}, \; a{:}{\color{violet} list}[{\color{violet} int}]) \rightarrow  {\color{violet} bool} : \\
   &
   {\color{cyan}\bf return}\  
   {\sf all}(retval \geq x \;{\bf for} \;x\; {\bf in}\; a) 
\end{array}
\] }

\HIDE{
{\scriptsize
\begin{lstlisting}[language=Python]
def post_M2(retval:int, a: list[int]) -> bool:
  if not a:
    return False
  return all(retval >= val for val in a)
\end{lstlisting}}
}

\noindent For the negative test $(20, [5, 2, 10])$, the post-condition {\sf post\_\!M2} evaluates to \emph{true}, therefore accepting an incorrect implementation of the {\sf findMax} program. 
We call this a {\bf false positive}, because {\sf post\_\!M2} incorrectly states that $20$ is the correct output for the input $[5, 2, 10]$. In this case, a buggy implementation of {\sf findMax} is accepted by the post-condition  {\sf post\_\!M2}.

The notions of \emph{false negative} and \emph{false positive} tests allow the pass/fail classification of non \emph{erroneous tests} to be refined. Consider the post-condition candidate $post_M$ generated by an LLM $M$ for a program $F$. The objective is to assess how accurately $post_M$ captures the correctness criteria defined by the oracle post-condition $post_F$. Evaluating $post_M$ against $post_F$ for a specific test $(o,x)$ yields four possible outcomes:
\begin{itemize}[leftmargin=*]
    \item \emph{True Positive} (TP): both $post_F(o, x)$ and $post_(o, x)$ are \emph{true}. This signifies that the LLM's post-condition correctly validates a correct program output.
    \item \emph{True Negative} (TN): both $post_F(o, x)$ and $post_(o, x)$ are \emph{false}.  In this case, the LLM's post-condition identifies an incorrect program output, thereby correctly flagging the program $F$ as faulty.
    \item \emph{False Negative} (FN): $post_F(o, x)$ is \emph{true}, but $post_M(o, x)$ is \emph{false}. This is a scenario where the LLM's post-condition is too strong, leading to the incorrect rejection of a valid program output. This could cause a correct program to be flagged as incorrect.
    \item \emph{False Positive} (FP): $post_F(o, x)$ is \emph{false}, but $post_M(o, x)$ is \emph{true}. This is a critical failure where the LLM's post-condition is too weak. Model $M$ incorrectly validates a wrong program output, thus failing to detect a bug and potentially leading to the acceptance of a faulty program. This is often considered a more dangerous error in verification than a false negative. 
\end{itemize} 

The False Negative Rate (FNR) \NEW{over the test cases} represents the proportion of positive cases that are incorrectly identified as negative. In other words, it reflects the likelihood of a correct program being flagged as incorrect. Conversely, the False Positive Rate (FPR) is the proportion of actual negatives that are incorrectly classified as positives. Put simply, it measures how often a post-condition leads to a positive result when the program is incorrect. More formally,
$$
    F\!N\!R = \frac{F\!N}{F\!N + T\!P} \quad \quad F\!P\!R = \frac{F\!P}{F\!P + T\!N} 
$$

The same considerations regarding test classification also apply to pre-conditions. 
%\todo{I think the following statement is not needed.} For a given a program $F$ and input $x$, the oracle pre-condition $pre_F(x)$ is \emph{true} if and only if $x$ is a valid input for $F$. Given the LLM generated pre-condition $pre_M$, the objective is to assess if $pre_M$ captures the correctness criteria defined by $pre_M$. \\

\NEW{FNR and FPR are further refinement of the aforementioned test passing rate metric. Recall that the latter gives an indication on how close an LLM solution is to being correct.
If a developer, hypothetically, adopts a post-condition from an LLM as is, and it turns out to be only near-correct, FNR gives an indication on how noisy the post-condition is by producing spurious bug alarms, and FPR is an indication of how lossy it is by failing to catch bugs.}\\

\noindent
{\bf Difficulty related metrics}

To evaluate the impact of task difficulty on LLMs' performance ({\bf RQ2}), we calculate the \emph{number of failing models per task}\NEW{, for each category of difficulty}. A model is considered to have failed 
a specific task if there is at least one failing test in each replica. \NEW{This could give us insight on, for example, whether across the board LLMs can handle pre- and post-conditions that involve quantifiers.}
Another important metric is the \emph{number of tasks \NEW{of each category of difficulty} with accept@\!1 > 0} for an LLM. If a model's accept@\!1 is not 0, this indicates that there is at least one replica in which the model proposes a correct solution for the task. \\

\noindent
{\bf Impact of automated testing}

Finally, we calculate the number of \emph{misclassified solutions} to assess the impact of automated generated tests on the benchmark. A solution $s$ proposed by a model $M$ for a task $T$ is misclassified if $s$ was deemed correct based on manual tests, but was found to be incorrect when executed against automatically generated tests. \NEW{The automatically generated tests thus manage to identify an incorrect solution, for which the manual test suite failed to do so. The number of misclassified solutions gives thus indication on the contribution of automatically generated tests in augmenting the manual tests in the dataset.}

\section{Results} \label{sec:results}
\begin{figure*}[tb]
    \centering
    \includegraphics[scale=1.1]{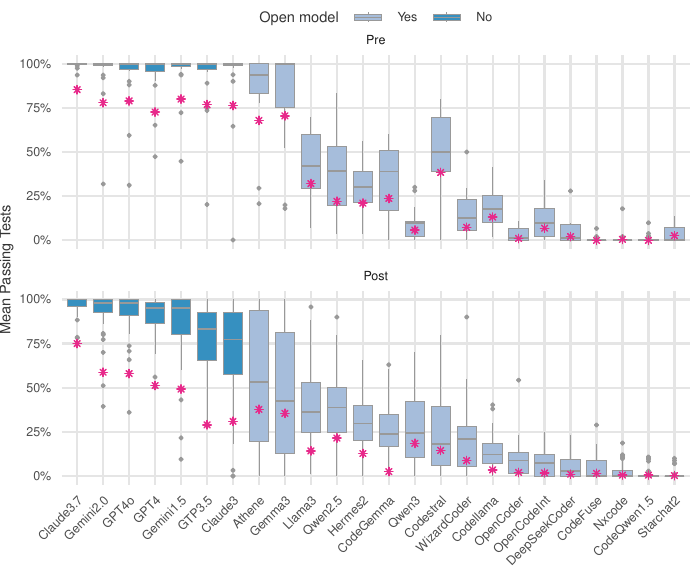}
    \caption{Boxplot showing the distribution of {\mpt}. Each dot represents the average across 10 replicas of successful tests for a task. The pink asterisk indicates the average accept@1 across all tasks. }
    \label{fig:averagePassBoxplot}
\end{figure*}

Here we present the results of the experiment and provide answers to the research questions. %detailed in the previous section. 

\subsection{RQ1 (LLM performance)}
RQ1 investigates the performance of LLMs in generating formal pre-/post-conditions from their natural language descriptions. Figure~\ref{fig:averagePassBoxplot} shows the results for all the LLMs considered in our experiment, based on the \mpt and \acceptone metrics (see Section~\ref{sec:metrics}). The boxplots show the values of the \mpt metric for each model, i.e., the proportion of the test cases that pass when executed on the code proposed by the LLM. The asterisks correspond to the \acceptone metric. The plot at the top shows the results for pre-condition generation and the plot at the bottom shows the results for post-condition generation, while the names of the models are shown on the x-axis, sorted in decreasing order of \mpt for post-conditions. Each boxplot in Figure~\ref{fig:averagePassBoxplot} corresponding to an LLM represents the results obtained when applying that LLM on all tasks (pre- or post-condition) each repeated 10 times. %The asterisk represents the \acceptone metric. %~\cite{endres2024can}.
%, which is the observed probability for the LLM to give a correct solution in a single attempt.
The boxplots corresponding to proprietary models are shown in blue colour, while those corresponding to open models are shown in light blue. Overall we can observe that the models, proprietary models in particular, tend to perform better on pre-condition generation than post-condition. We can also observe that proprietary models perform generally better than the open models, across all tasks, in terms of both the \acceptone and mean passing tests metrics.
\NEW{We can further notice that the proprietary models have high mean passing tests numbers, considerably higher than their \acceptone. This suggests that their solutions, even if not correct, are still often near correct. Section \ref{sec:discussion} will show some examples of these.}

\begin{figure*}[htb]
    \centering
    \includegraphics[scale=1.1]{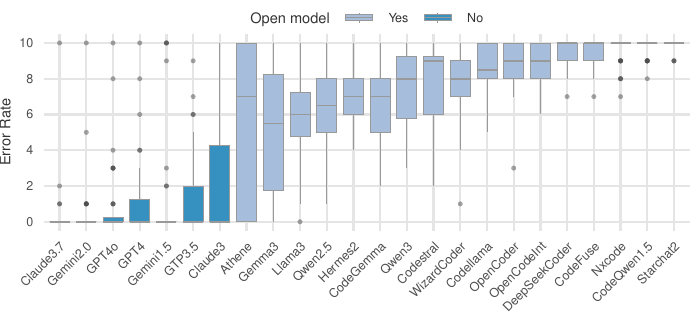}
    \caption{Number of replicas where the LLM proposed solution on a task is not valid because at least one
    test returns an \emph{error} when executed on it. %\XCOMMENT{maybe also add 'Each dot represents ...' as in Fig-4.}}    
    \NEW{Each dot represents a task. For instance, the dot with value ten for Claude3.7 model means that there is a task for which all ten replicas contain an execution of a test that results in an \emph{error}.}}  
    \label{fig:averageErrorBoxplot}
\end{figure*}

The plots shown in Figure~\ref{fig:averagePassBoxplot} are based on pre-/post-conditions returned by the LLMs and were parsed correctly as executable Python code, i.e., execution of the generated code gives pass/fail result. However, in our experiments we have noticed the presence of quite a significant proportion of pre-/post-condition code returned by LLMs that results in \emph{error} where the returned code either does not represent a valid Python code or is somehow malformed and/or returns values outside of the expected range, thus resulting in an error. Figure~\ref{fig:averageErrorBoxplot} shows, for each task in the experiment, the number of runs (out of 10) that resulted in \emph{error} for each of the LLMs, i.e., the \errorrate. Each dot in the plot represents the number of times an LLM gave an erroneous response for a task (pre-/post-condition generation) out of 10 replicas. We consider a solution proposed by an LLM to be erroneous if the execution of at least one test case results in an erroneous output. Python being an interpreted language, some of the errors are revealed only in certain execution scenarios, which in our experiments are represented by different test cases, both written manually and generated automatically (see Section~\ref{sec:experiment-design} for details). For example, on the ${\sf HEx}_{57}$ task, the best performing model (Claude3.7), out of the 10 replicas, it gave 9 correct results and one erroneous result. The erroneous result causes a runtime Python error when executed on a test case. % \todo{mention HE57?}

\begin{figure*}[htb]
    \centering
    \includegraphics[scale=1.1]{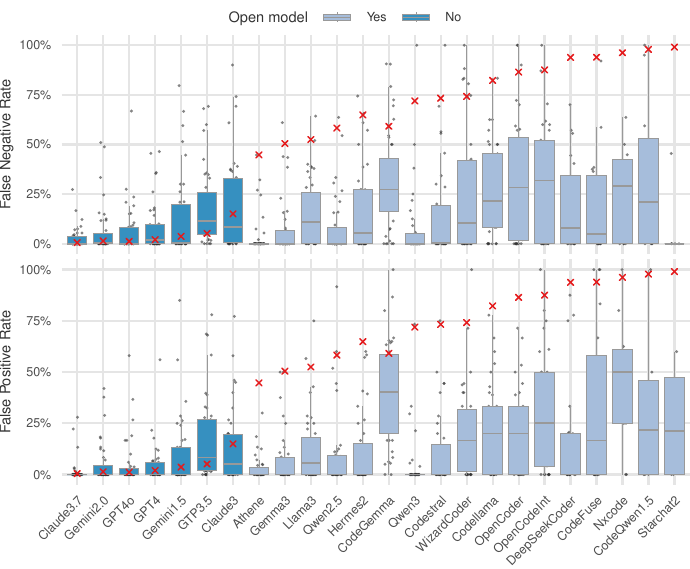}
    \caption{Boxplot showing the False Negative (above) and Positive (below) Rates per task. Each dot shows the average FNR/FPR across ten replicas of a model on a task. Red cross mark indicates the \errorrate.}
\label{fig:FPRndFNRpostBoxplot}
\end{figure*}

As can be seen from Figure~\ref{fig:averageErrorBoxplot}, the open LLMs tend to produce a high number of code that ends up in error. A closer inspection 
%of the output produced by the LLMs 
reveals that the \emph{error} cases are mainly due to: 1) the LLM failing to produce correct Python code, 2) the pre-/post-condition function returns an unexpected value, e.g., an integer instead of boolean, 3) the LLM does not follow the prompted instruction well e.g., by not producing any code at all, or by returning a lengthy explanation text from which it is difficult to reliably extract the code part. 
Note that the prompt in Figure~\ref{fig:prompt} explicitly instructs the LLM \emph{not} to provide an explanation.
%\XCOMMENT{There is another reason: the response from the LLM is not in the right format. Open LLMs may talk too much :) , and it becomes problematic to extract the code from the answer. Note that our prompt instructs them to only give the  code (without explanation). Propiatry LLMs do understand this instruction. I'll add this, extending 3) above}

% \todo{detail better these cases, maybe give a proportion?} 

%\todo{@Davide: update the y axis label of Figure~\ref{fig:averageErrorBoxplot} to "Replicas with an error"}

The performance results shown in Figure~\ref{fig:averagePassBoxplot} should be interpreted in light of the \errorrate presented in Figure~\ref{fig:averageErrorBoxplot} where we can observe the general pattern that the best performing models tend to be those with lower \errorrate values. It is also interesting to note, however, that even the best performing models produce erroneous proposals, as demonstrated by the presence of outliers in the plot shown in Figure~\ref{fig:averageErrorBoxplot}.

% \begin{figure}[tb]
%     \centering
%     %\includegraphics[width=1\linewidth]{figures/fp_and_fn_post_boxplot.pdf}
%     \includegraphics{figures/fp_and_fn_post_boxplot.pdf}https://www.overleaf.com/project/6761db8ec3d8ca10fe40594f#
%     \caption{Percentage of FP (above) and FN (below) tests per task. Each dot represents the average FP/FP across 10 replicas of a model on a task.}
% \label{fig:FPndFNpostBoxplot}
% \end{figure}

%\todo[inline]{Need to update the text accounting for False Positive and Negative Rates}

Delving further into the results, we inspected the cases where the models gave the wrong answer, i.e., the proposals generated by the models fail on one or more tests. Please note that these cases do not include the erroneous cases discussed in the previous paragraph. To get insight into the nature of the wrong proposals produced by the models, we inspect the data related to cases of \fp and \fn (see Section~\ref{sec:metrics}). 
% categorized them into \emph{false positive} (FP) and \emph{false negative} (FN). Given a test case, a FP outcome for the test is one where the model generated contract gives a true output, while the expected output should be false. Similarly a FN outcome for a test is one where the model generated output is false but the expected output is true. 
Figure~\ref{fig:FPRndFNRpostBoxplot} shows the distribution of \fnr (FNR, top plot) and \fpr (FPR, bottom plot)  for each model on the post-condition generation tasks. 
\NEW{Recall that FNR is an indicator of how noisy a proposed pre-/post-condition is, whereas FPR indicates how lossy it is.}
%More specifically, the plots show the proportion of tests that give \fp and \fn outcomes. 
As can be seen from the plot in Figure~\ref{fig:FPRndFNRpostBoxplot}, the LLMs generally have a low FNR, i.e., the post-condition proposed by the LLM has a low probability of incorrectly marking a correct program as wrong. 
Some open source models, such as Starchat2 and Athene, appear to have a lower FNR compared to closed models. However, when comparing different LLMs, the \errorrate should be considered  (red cross mark in Figure~\ref{fig:FPRndFNRpostBoxplot}). While it is true that Starchat2 has a very low FNR, it was only able to produce a non-error solution on four tasks. The same applies to FPR. 
\NEW{Some closed models have quite low FNR and FPR as well as Error Rate. Even a higher FNR may not be bad, as a spurious error alarm would trigger the developer to investigate and fix the pre- or post-condition he/she adopted. Once fixed, the error will not appear again. 
On the other hand, low FPR still means that there is risk. It is then prudent to complement an LLM-generated pre- or post-condition, if adopted, with e.g. manual tests.}
When we consider closed models, we can also see the improvements in terms of lower FNR and FPR across different versions of the same model.  For instance, when considering FPR, Claude3 has a median FPR of 5\%, which is reduced to zero in the new Claude3.7 version. However, even the top-performing models show some outliers, with peaks exceeding 25\% for both FNR and FPR. The results for pre-condition tasks are generally similar to those of post-condition tasks but the FPR and FNR values tend to be lower. The detailed results are included in our replication package.   \\

%the proportion of tests resulting in \fp/\fn outcomes is generally low, more so for \fp than \fn. Interestingly, we observe that the proprietary models tend to have higher proportions of {\fn}s, even though the median values are low. It should also be mentioned here that the plots show the percentage of \fp/\fn tests, and as discussed earlier (Figure~\ref{fig:averageErrorBoxplot}), the non-proprietary models produce a disproportionately high number of proposals that end up with an \emph{error} outcome.\\

\hspace{-4mm}
\begin{minipage}[c]{0.97\columnwidth}
\begin{tcolorbox}[colback=white]
Overall, the LLMs were able to generate valid pre- and post-conditions for the majority of the tasks, with the proprietary models being significantly more effective than open models. Looking at the wrong solutions, \NEW{those from proprietary models are often near-correct, with high tests passing rate.} FPR and FNR are generally low on pre-condition tasks than on post-condition tasks. Proprietary models have slightly higher FPR and FNR on post-condition tasks than on pre-conditions. Results also show that open models generate an overwhelmingly large number of erroneous solutions.
\end{tcolorbox}
\end{minipage}

\subsection{RQ2 (Impact of task difficulty on LLM  performance)}

\begin{figure*}[tb]
    \centering
    \includegraphics[scale=1.2]{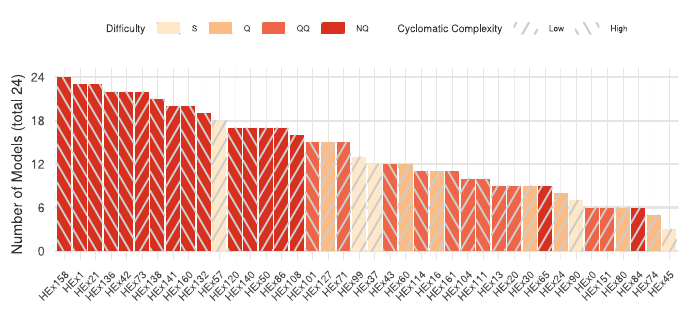}
    \caption{Number of models failing to produce a valid post-condition for each task \NEW{{\bf across different difficulties}}. Colours indicate the difficulty category of a task, and  strips the cyclomatic complexity.}
    \label{fig:difAndCcBarplot}
\end{figure*}

% \begin{figure}[htb]
%     \centering
%     \includegraphics[width=1\linewidth]{figures/cc_barplot.pdf}
%     \caption{The barplot shows the number of models failing to produce a valid post-condition for each task, with colors indicating the post-condition cyclomatic complexity.}
%     \label{fig:ccBarplot}
% \end{figure}
With {\bf RQ2} we investigate the performance of the LLMs as a function of the difficulty of the task. We consider two notions of difficulty, i.e., cyclomatic and semantic complexity. The details of these metrics and how they are computed is discussed in Section~\ref{sec.spec.diff}, while a summary of the complexity metrics for our dataset is presented in Table~\ref{tab:HEx}.
\subsection{RQ2.1 Cyclomatic complexity}
As shown in Table~\ref{tab:HEx}, the tasks in our dataset are grouped into \emph{low} (CC $\leq$ 5) and \emph{high} (CC > 5) complexity based on the cyclomatic complexity of the (manually written) functions that implement the contracts (Section~\ref{sec:prepost}).
Figure~\ref{fig:difAndCcBarplot} shows, for each post-condition 
%\footnote{here we consider only post-condition tasks as they are the most complex, while pre-conditions are generally not so complex} 
generation task, the number of LLMs that fail to produce a valid post-condition. Notice that in the plot in Figure~\ref{fig:difAndCcBarplot} the cyclomatic complexity of each task is represented by the patterns of the stripes. As can be seen from the figure, the tasks with high cyclomatic complexity indeed challenge the LLMs where in the extreme case (${\sf HEx}_{158}$) none of the LLMs were able to generate a valid solution for it. For the tasks with low complexity the number of LLMs that fail to generate a valid solution drops considerably. There are however cases where the complexity class does not appear to match the observed results, i.e., low complexity tasks leading to a large number of LLMs failing to generate a valid solution (e.g., ${\sf HEx}_{73}$, ${\sf HEx}_{141}$, ${\sf HEx}_{57}$). On the contrary, some of the high complexity tasks do not appear to be as challenging (e.g., ${\sf HEx}_{111}$, ${\sf HEx}_{161}$, and ${\sf HEx}_{114}$). Some of these outlier cases are discussed in greater detail in Section~\ref{sec:discussion}. Overall, we can observe that tasks with high cyclomatic complexity do pose challenges to the LLMs' capability of generating valid post-conditions.

\subsection{RQ2.2 Semantic complexity}

\begin{figure*}[tb]
    \centering
    \includegraphics[scale=1.2]{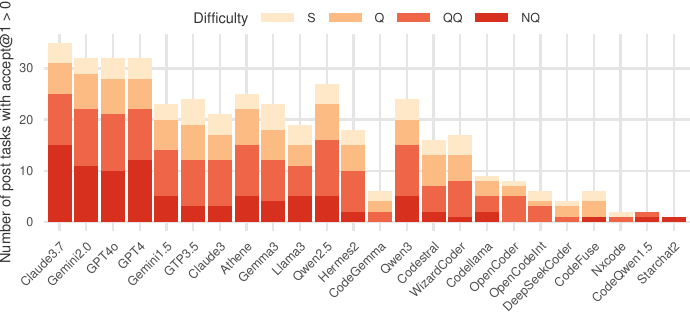}
    \caption{Number of correct post-conditions\NEW{, {\bf across different difficulties},} generated by each model. }
    \label{fig:difficultyBarplotByModel}
\end{figure*}

% \begin{figure}[htb]
%     \centering
%     \includegraphics[width=1\linewidth]{figures/difficulty_barplot.pdf}
%     \caption{The barplot shows the number of models failing to produce a valid post-condition for each task, with colors indicating the difficulty category.}
%     \label{fig:difficultyBarplot}
% \end{figure}
While cyclomatic complexity is an indirect measure of a task's difficulty, as it is based on the complexity of the function written to implement it, in RQ2.2 we look at a measure of complexity of the task considering its natural language specification. Our analysis (see Section~\ref{sec:prepost}) led to four complexity classes to which the tasks in our dataset belong (see Table~\ref{tab:HEx} for a summary). Similarly to RQ2.1, we consider post-condition tasks only. We can see from Figure~\ref{fig:difAndCcBarplot} (colored shades) \NEW{that generally more than half of the tested LLMs are able to produce a correct solution for tasks with difficulty S, Q, or even QQ, indicating that these are the kind of tasks that are reasonably within the current LLMs capability.}
However, tasks classified as NQ (non quantifiable) do indeed challenge the LLMs leading to a majority of the LLMs not producing valid post-conditions. With the exception of a few cases (e.g., ${\sf HEx}_{57}$) the tasks on which a large number of LLMs fail to produce valid solutions belong to the NQ class. Aside from a few outliers where few models were challenged by NQ tasks (${\sf HEx}_{65}$ and ${\sf HEx}_{84}$), overall the difficulty class, similar to the cyclomatic complexity, highlights the challenge faced by LLMs when dealing with difficult specifications.

To get further insight, we also analysed, for each LLM, how many of the tasks of each difficulty class it was able to find a valid solution for. In Figure~\ref{fig:difficultyBarplotByModel} we plot the number of post-conditions each model was able to find at least one correct solution out of the 10 replicas, distinguished by the difficulty class. We observe that
\NEW{proprietary models and some of the open models (e.g. Qwen) are able to handle S, Q, and QQ tasks well.}
The newer proprietary models  perform well on the difficult tasks 
%(QQ and NQ) 
(NQ)
while the older ones (Gemini1.5, GPT3.5) do not perform as well as their latest variants. It is also worth noting that the number of low-difficulty tasks (S and Q) tends to be stable over a large number of the LLMs. Though we do not show the plot here due to space limitations, a similar trend is observed for cyclomatic complexity as well. \\

\hspace{-4mm}
\begin{minipage}[c]{0.97\columnwidth}
\begin{tcolorbox}[colback=white]
Our dataset is composed of tasks with varying 
%levels of 
difficulty, and 
%\The results of 
our experiment reveals that the performance of LLMs is significantly affected by the difficulty level of the task at hand, 
\NEW{performing well on non-NQ tasks, but not on NQ ones}.
In the extreme case, none of the 24 LLMs in our study were able to generate a valid solution for one of the tasks. On the other hand, the best performing models are able to solve a higher number of difficult tasks.
\end{tcolorbox}
\end{minipage}

\vspace{3mm}
\subsection{Impact of automatically generated tests}
% What is the impact of adding automatic tests? Do automatic tests change the evaluation of a proposed solution, i.e., a solution considered valid using only manual test can become invalid adding automatic tests? 

\begin{figure*}[tb]
    \centering
    \includegraphics[scale=1.2]{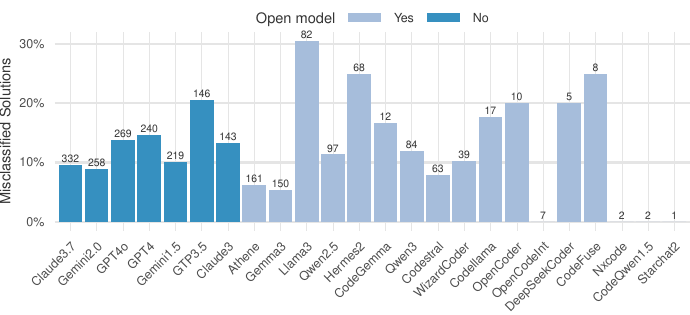}
    \caption{Percentage of invalid responses for post-conditions when considering only manual tests. The numbers above the bars show the total number of valid solutions when using manual tests only.}
    \label{fig:misclassifiedBarplot}
\end{figure*}

The results discussed so far are based on the entire set of test cases in our dataset, i.e., both tests written manually and those generated automatically (Section~\ref{sec:methodology}). Here we take a closer look at the results in order to get insight into the role of the tests that were automatically generated. For this purpose we analyzed the results by distinguishing the cases where the LLMs produced correct results when considering \emph{only manually written tests} but the outcome changed (i.e., the solution proposed by the LLM turned out to be incorrect) when considering automatically generated tests as well. In other words, we computed for each LLM how many proposed solutions would have been considered correct if we had used only the manually written tests for checking the correctness. For example, when considering only manual tests, 9.6\% (32) of the valid solutions  proposed by Claude 3.7 are actually invalid when automatically generated tests are also used. Keep in mind that the manually written tests achieve full branch coverage of the code.  In Figure~\ref{fig:misclassifiedBarplot} we plot the number of  cases where the use of automated tests exposed a wrong solution, i.e., solutions that were deemed correct with respect to manual tests, but turned out to be wrong when executed against automatically generated tests. As can be seen from the figure, the automatically generated tests exposed several misclassified cases.  It is worth mentioning that the open LLMs (the right half of the figure) generate very few valid (non erroneous) solutions, as shown in Figure~\ref{fig:averageErrorBoxplot}, hence the plot of misclassified solutions is based on a small number of cases. %As a result, the proportion of misclassified solutions swings radically between high values (e.g., 1 out of 2 would be 50\%) and low values (e.g., 0 out of 2 would be 0\%). On the other hand, the proprietary models produce very few erroneous solutions, hence the values in the plot are statistically more stable. %\todo{refine the numbers} 

\section{Discussion} \label{sec:discussion}
\HIDE{
In this section we discuss in detail some selected cases from our experimental study and insight learned from them.
}
%that exposed different weaknesses in the different LLMs.
%and the contribution of automatically generated test cases in exposing wrong solutions proposed by LLMs which would otherwise have been accepted as correct solutions.
%\subsection{Place holder}

\subsection*{Erroneous solutions from open LLMs}

\begin{figure*}[htbp]
    \centering
    \includegraphics[scale=1.1]{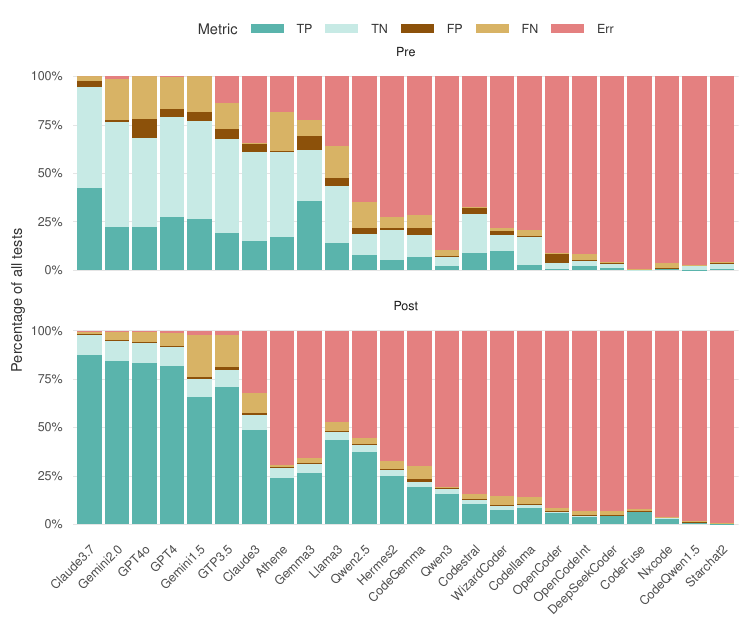}
    \caption{The barplot depicts the execution output of all executed tests for each model for pre- and post-conditions.  The test results are categorised as true positive (TP), true negative (TN), false positive (FP), false negative (FN) and error (Err), in accordance with the metrics defined in Section~\ref{sec:metrics}.  }
    \label{fig:testOutData}
\end{figure*}
As discussed in Section~\ref{sec:results}, open LLMs tend to produce more erroneous solutions. \NEW{To get a more comprehensive view of the successes and failures of the various LLMs, we have plotted the outcomes from all executions and all tasks in Figure~\ref{fig:testOutData}. As we can see from the figure, \emph{erroneous} solutions account for a large proportion of the failures observed for the open models, where the proportion of \emph{wrong} solutions (false positives/negatives) are much less than the proportion of \emph{erroneous} solutions.}
One of the reasons is that some LLMs verbosely respond to prompts by generating explanations. This is the case with the five LLMs with the highest number of erroneous proposals. %(Figure~\ref{fig:averageErrorBoxplot}). 
These are also the lowest performing LLMs (Figure~\ref{fig:averagePassBoxplot}).
Note that the prompt in Figure~\ref{fig:prompt} explicitly asks the LLMs {\em not} to explain, as we are only interested in the code. These LLMs fail to understand the instruction well. For instance, Nxcode responds with an  explanation but does not give the code. Deepseek-coder typically gives lengthy explanations and the code is embedded somewhere in the explanation. It then becomes problematic to reliably extract the code from the surrounding text. In contrast, the top performing LLMs clearly understand the instruction and return only the code; and if explanation is provided, it is properly included as code comments.
What this tells us is that just training an LLM with programming skill is not sufficient. To be useful for end users, it also needs to be trained to follow instructions (so-called {\em instruct-training}). Propriety LLMs perform well on this, whereas the open source community still has some gap to close on this aspect.

\subsection*{Difficult and subtle cases} \label{sec:subtle-cases}

The post-condition ${\sf HEx}_{104}$ mentioned in Section~\ref{sec:dataset}, which is of difficulty $\sf QQ$, is non trivial. Figure~\ref{fig:difAndCcBarplot} shows that most LLMs can actually solve it. 
Furthermore, it turns out that seven open LLMs can solve it (not shown in the figure). This is certainly good news for open LLMs.
If we however look at the
ten most challenging post-conditions in Figure~\ref{fig:difAndCcBarplot} (left most ten columns), they are all of the difficulty $\sf NQ$, which is perhaps not surprising. The good thing is that all of them, except ${\sf HEx}_{158}$, are at least {\em collectively} solvable by the LLMs
(that is, for each of these post-conditions, at least one LLM manages to produce one good solution within the allotted ten tries). 
Unfortunately though, with the exception of ${\sf HEx}_{21}$ and ${\sf HEx}_{141}$, none of the solving LLMs are open LLMs. This emphasizes the previously made observation on the significant gap between the open and proprietary LLMs. 
  
The open LLM Llama3 (70B) is the only one that can produce a correct solution for the post-condition of ${\sf HEx}_{21}$. However, GPT4o also came up with a solution that passed all the tests, but one. ${\sf HEx}_{21}$ involves calculations such as $\lfloor a*b/c \rfloor$ that can be sensitive to the imprecision of $\sf float$ arithmetic. The failing test exposes the imprecision. If this is tolerated, then arguably GPT4o's solution can be considered as correct as well. Neither solution employs robust floating point arithmetic (e.g. using Python's $\sf Decimal$ datatype), but to be fair, the  natural language description of ${\sf HEx}_{21}$ 
 did not explicitly require that either.

${\sf HEx}_{158}$ is an interesting case as no LLM can solve it. The post-condition is of type $\sf NQ$. Its description:
%, along with an example:\\

{\small
\begin{quote}
Given a list $s$ of strings, the program returns a string with the {\bf longest stutter}. A stutter of a string is a segment of length at least two, consisting of the same letter. If no string in $s$ has a stutter, the program returns $\sf None$. \\
{\bf Example:} 
{\tt \scriptsize
$\sf Pr\_HE158$(["hiii!","woorld"]) = "hiii!"}\\
\end{quote}
}

Claude-3.7 and Gemini-2.0 produce solutions that pass 93\% of the tests (so, near correct), failing on three tests that check that only repeated {\em letters} count as a stutter, in accordance with the description. Hence, "ii" counts as a stutter, but "11" does not. This subtle point is missed by both Claude-3.7 and Gemini-2.0. To be fair, a human programmer may miss the subtlety as well.

Another subtle case is the post-condition of  $\sf HEx_{50}$:

{\small
\begin{quote}
Given a string $s$, the program returns an encoded version of s. This is
obtained by shifting every character $c$ in $s$.
A letter is cyclicly shifted by $k$ symbols in the alphabet. We keep the case (e.g.
if it was lower case, it remains lower case). Digits in a number are shifted
$k$ symbols in 0..9.
Non alpha-numneric characters, except spaces, are left unchanged.
Segments consisting of only {\color{red}spaces are replaced by a single dash}.
\\
{\bf Example:} {\tt\scriptsize $\sf Pr\_HE50$("h3i!",1) = "i4j!"}
\end{quote}
}
\noindent
The post-condition, which is of $\sf NQ$ difficulty, is solved by six LLMs (the LLMs give at least one correct proposal). Interestingly, some wrong solutions from Claude-3.7 failed {\em only on one test}. A simplified version of this test gives  
$s = "{\backslash}x0b"$ and $k{=}1$ as the input, and poses the question whether the same $s$ would be a correct output $t$ of  the program $\sf Pr\_HE50$. 
In Python, ${\backslash}x0b$ is a code for the vertical tab character.  
A correct post-condition should approve this $t$. A vertical tab is not a letter, and technically it is not a space either. So, it should not be shifted.
The wrong post-conditions rejected the $t$, insisting that the correct $t$ should be a "-", as per the red part of $\sf HEx_{50}$'s description shown above. These post-conditions consider all {\em white space} characters, including a vertical tab, as spaces. Indeed, arguably, the given natural language description was ambiguous.

So, just like humans, LLMs are  prone to natural language pitfalls e.g. implicit information (${\sf HEx}_{21}$), ambiguity (${\sf HEx}_{50}$), and nuances (${\sf HEx}_{158}$). Using an LLM still of course saves us from spending human effort.
%, and in all the above cases, the LLM either manages to produce at least one good answer, or near good ones.
One may speculate whether a solution with high test passing rate can be considered as 'almost good', and is therefore valuable. For the above three cases, the 'almost good' LLM solutions are actually quite well readable. Given the failing test cases, an experienced programmer can easily fix them. 
One might then wonder: 
how hard would it be, more generally, to fix 'almost good' LLM solutions? This would make an interesting future work.

It is also worth noting that the discriminating tests mentioned in ${\sf HEx}_{21}$ and ${\sf HEx}_{50}$ were generated by Pynguin. This suggests that augmentation with automated testing has a key role to guard against LLMs susceptibility to language subtleties. Human testers are prone to that as well, while a tool like Pyguin is not.

\subsection*{Should we show the program's code?}

Our prompt (Figure~\ref{fig:prompt}) does not send the program's code, it only shows the program's natural language description.
This is a deliberate choice as we want to focus on contract generation purely from natural language descriptions. Also, as remarked in Section~\ref{sec:releated-work},
including the program code may pass on bugs in the code to the produced contract.
It is also unclear whether including code in the prompt would actually help the LLM. In the study by \cite{endres2024can} this does not lead to a significant improvement; in some cases even worsens the performance. On the other hand, tools like AutoSpec and SpecGen only use the program code (without program description) and showed that LLMs can generate useful assertions, despite being partial. \NEWB{\cite{faria2026automatic} use both code and descriptions inside the code in an iterative approach with promising results.} This could be further investigated as future work, exploring different ways of combining description-based and code-based  approaches.

\NEWB{
\subsection*{What does this mean in practice?}

As mentioned in the conclusion to RQ1, as it is now, the LLMs, in particular the high end proprietary models, are able to produce correct or near correct full pre- and post-conditions in our dataset, indicating that adopting them in actual practice is a very plausible idea, at least for unit-level programs.
Also, one can expect LLMs to become better in the future, e.g. due to increase in size and intensified training. 
We cannot however expect LLMs to be flawless, both due to hallucination that is inherent in them and due to ambiguity and incompleteness inherent in natural language descriptions, leading to subtle cases as discussed earlier in Section~\ref{sec:subtle-cases}. This means that employing an LLM-based generation in practice requires a good test suite as a guard rail, enabling developers to detect and fix subtle errors. Writing tests has of course its own cost, so we would want to know how to reduce this cost while still getting good quality pre- and post-conditions. This is as far as we know still an open question and would make an interesting future work.
}

\section{Threats to Validity } \label{sec:threats}
% Internal T:

% \begin{itemize}
%     \item The tests might not be thorough enough. Mitigation: test augmentation.
%     \item Solution pre- and post are incorrect. Mitigation: manually inspected by an expert. Pre-conds are typically simple, so easy to check. Post-conditions are cross checked with the provided implementation of the PUT, via testing.
%     \item Programming error in calculating metrics. Mitigation: two implementations of the calculation, cross checked.
% \end{itemize}

% number of models
% programming language
% correctness of manually written solutions
% leakage of the dataset
% non determinism of LLMs and the number of repetition
% validation of tests: 
% manual check
% for post, we give the expected program output in the test but we also check against real program output (by definition in automated testing)

There are potential threats to the internal and external validity of our results. Internal threats could arise from our dataset where pre- and post-condition could be incorrect, as they are prepared manually by us. Pre-conditions are typically simple and easy to check. Post-conditions are checked via testing. Still, potential error cannot be altogether excluded. Another threat is the fact that the test cases in our dataset may not be adequate for checking the correctness of the generated contracts. To mitigate this threat we have augmented the manual tests through automated test generation, which indeed proved to be fundamental. We have used a newly constructed dataset with the objective of avoiding potential leakage from LLMs having already seen the dataset. Yet another threat is the non deterministic responses from LLMs. We have repeated every LLM prompting 10 times to account for this aspect, however a larger number of repetitions may be necessary for a more reliable results. Regarding external validity, we have used 24 proprietary and open models which are currently in use both in research and practice, however a larger number could increase confidence in the results. Finally, in our experiment, the contracts generated are in Python, and results may be different for other programming languages.

\section{Conclusion and Future Work}\label{sec:concl}

We have evaluated 24 LLMs, seven of which are proprietary, on the task of formalizing contracts ({\em full} pre- and post-conditions) of programs, given their natural language descriptions. A new dataset, called $\sf HEx$, was used. On average, the proprietary LLMs performed significantly better than the open LLMs, with the top five models being all proprietary.
At the other end, the five lowest performing LLMs are all open models. These LLMs
produce a` high number of erroneous/ill-formed proposals.
This is not necessarily due to lower programming skill. We observed that these LLMs struggle in interpreting the instructions in the prompt. A better instruct-training may significantly improve their performances.

%Claude3.7 gave the best performance of  0.85 average accept@\!1 over combined pre- and post-conditions of all problems in the dataset.
%The best performing open model is Gemma3, with 0.71 average accept@\!1, close to GPT4's performance.
%Interestingly, Gemma3 'only' has 27B parameters, compared to e.g. Athene or Llama3 with 70B parameters.

Claude3.7 gave the best performance of  0.78 average accept@\!1 over combined pre- and post-conditions of all problems in the dataset.
The best performing open models are Gemma3 and Athene, with 0.47 average accept@\!1, beating Claude3 and GPT3.5 proprietary models.
Interestingly, Gemma3 'only' has 27B parameters, compared to e.g. Athene or Llama3 with 70B parameters.

Generally, the LLMs perform better on pre-condition tasks. This is expected, as pre-conditions are typically less complex.
None of the 24 LLMs can solve all pre- and post-condition tasks (within the tried ten attempts).
In particular,  generating correct contracts of difficulty $\sf NQ$ is challenging for the LLMs.
The good news is that the best performing LLMs are able to solve a higher number of $\sf NQ$ tasks.
%However, the good news is that collectively they can solve all tasks, including the $\sf NQ$s, except the post-condition of ${\sf HEx}_{158}$.
The relatively low proportions of false positives and negatives suggest that proposed solutions, while incorrect, might  be not too far from being correct. In some cases, they indeed can be easily fixed by a human programmer. Whether this would hold more generally is beyond the scope of the current study. This would be an interesting future work.

A good test suite is needed to validate LLM-generated contracts. Tests written by the developers themselves, even if they have full code coverage, while certainly useful, are also often biased. Augmenting human tests with automatically generated tests were observed to be critical in rejecting wrong contracts that would have otherwise been accepted.

Closer inspections of selected cases reveal that LLMs are susceptible to language subtleties such as implicitness, ambiguity, and nuances  in the human-written program description, which is the source of their inference.
They can propose contracts that look deceptively correct, but are actually wrong. Again here, automated testing, which carry no human bias, proved to be useful in differentiating such proposals.

For future research, we would like to look into specification generation for larger program units, e.g. classes.
Also, it is important to look into mechanisms to improve incorrect proposals.
We also believe that the sensitivity towards language subtleties needs more in depth studies.

\ \\

\ \\

\noindent {\bf Data:} a package with our data is available via Zenodo \url{https://zenodo.org/records/17499683}.

%
% By masking the part suspected to need repeair, by traning separate LLMs specialized in repair
%

% \noindent
% {\bf Acknowledgement:} we thank \censor{Morgaine Prasetya} for her initial work on the evaluation framework. We thank \censor{something something} for his help in test augmentation.

% \HIDE{
% \subsection*{\textbf{Replication package availability:} a replication package with all data and scripts is available via Zenodo \url{https://zenodo.org/records/17499683} as supplementary material.
% %It will be made publicly available upon paper acceptance.
% } }

%\clearpage

%\section{Declarations}
%\input{sections/declarations}

%\clearpage

%\printcredits

%% Loading bibliography style file
%\bibliographystyle{model1-num-names}
\bibliographystyle{cas-model2-names}
% Loading bibliography database
\bibliography{IEEEabrv,llms}

%\vskip3pt

%\bio{}
%Author biography without author photo.
%\endbio

%\bio{figs/cas-pic1}
%Author biography with author photo.
%Author biography. Author biography. Author biography.
%\endbio

\end{document}